\documentclass[useAMS]{mn2e}
\input{psfig}
\usepackage{times}
\usepackage{bm}
\usepackage{amsmath}  
\usepackage{url}

\def\Real{{\rm I\mathchoice{\kern-0.70mm}{\kern-0.70mm}{\kern-0.65mm}%
  {\kern-0.50mm}R}}

\font \bolditalics = cmmib10
\def\bx#1{\leavevmode\thinspace\hbox{vrule\vtop{\vbox{\hrule\kern1pt
        \hbox{\vphantom{\tt/}\thinspace{\bf#1}\thinspace}}
      \kern1pt\hrule}\vrule}\thinspace}

\def \vc #1{{\textfont1=\bolditalics \hbox{$\bf#1$}}}

\newcommand{\bea}{\begin{eqnarray}}
\newcommand{\eea}{\end{eqnarray}}
\newcommand{\vectii}[2]{\rund{\begin{array}{c} #1 \\ #2 \end{array} }}
\newcommand{\Om}{\Omega_\mathrm{m}}
\newcommand{\map}{M_\mathrm{ap}^2}

\def\be{\begin{equation}}
\def\ee{\end{equation}}

\def\ba{\begin{eqnarray}}
\def\ea{\end{eqnarray}}
\def\vp{\varphi}
\def\vt{{\vartheta}}
\def\Z{{\cal Z}}

\def \vc #1{{\textfont1=\bolditalics \hbox{$\bf#1$}}}{\catcode`\@=11
\def\eck#1{\left\lbrack #1 \right\rbrack}

\def\rund#1{\left( #1 \right)}
\def\abs#1{\left\vert #1 \right\vert}

\def\ave#1{\left\langle #1 \right\rangle}
\def\eps{{\epsilon}}

\def\d{{\rm d}}

\def\ltsima{$\; \buildrel < \over \sim \;$}
\def\lsim{\lower.5ex\hbox{\ltsima}}
\def\gtsima{$\; \buildrel > \over \sim \;$}
\def\gsim{\lower.5ex\hbox{\gtsima}}

\begin{document}
   \title[]{Sources of contamination to weak lensing tomography: redshift-dependent shear measurement bias}
\author[]
{\parbox[]{6.in}{Elisabetta
    Semboloni$^{1},$\thanks{sembolon@astro.uni-bonn.de} Ismael Tereno$^{1}$,
     Ludovic van Waerbeke$^{2}$, Catherine Heymans$^{2,3}$\\
  \footnotesize $^{1}$ Argelander-Institut f\"ur Astronomie, Auf dem H\"ugel 71,  Bonn, D-53121, Germany. \\
  \footnotesize $^{2}$  University of British Columbia, Department of
     Physics \& Astronomy, 6224, Agricultural Road, Vancouver, B.~C., V6T,
     Canada. \\
 \footnotesize $^{3}$ Institute for Astronomy, University of Edinburgh, Royal
     Observatory, Blackford Hill, Edinburgh, EH9 3HJ, UK. \\ 
}}
\maketitle
\begin{abstract}
The current methods available to estimate gravitational shear from astronomical images
of galaxies introduce systematic errors which can affect the accuracy of weak lensing
cosmological constraints. We study the impact of KSB shape measurement bias on the cosmological
interpretation of tomographic two-point weak lensing shear statistics.

We use a set of realistic image simulations produced by the STEP collaboration to
derive  shape measurement bias as a function of redshift. We define
biased two-point weak lensing statistics and perform a likelihood analysis for
two fiducial surveys. We present a derivation of the
covariance matrix for tomography in  real space and a fitting formula to
calibrate it for non-Gaussianity.

We find the biased aperture mass dispersion is reduced by  $\sim 20\%$ at 
redshift $\sim 1$, and has a shallower scaling with redshift. 
This effect, if ignored in data analyses, biases $\sigma_8$ and $w_0$ estimates by a few percent.
The power of tomography is significantly reduced when 
marginalising over a range of realistic shape measurement biases. For a
CFHTLS-Wide-like survey, $[\Om,\sigma_8]$ confidence regions are degraded by a
factor of 2, whereas for a KIDS-like survey the factor is 3.5. 
Our results  are strictly valid only for  KSB
methods but  they demonstrate the need to marginalise over a redshift-dependent
shape measurement bias in all future cosmological analyses.
\end{abstract}


\section{Introduction}
Weak gravitational lensing by large-scale structure, or cosmic shear (see Munshi et al. 2008 for a review), is a powerful tool to
investigate dark energy and the large-scale distribution of dark matter
\cite{detf,esoesa}. In the last decade,
two-point shear statistics have been successfully measured and used to constrain parameters
of the matter power spectrum (from the early constraints of van Waerbeke et
al. 2001 to the recent results of Fu et al. 2008) and dark energy
\cite{Jaetal06,Hoetal06,Seetal06,Schetal07,Ketal08}. 

The information contained in the cosmic shear signal is more efficiently exploited if the redshift of the source galaxies is available.
In particular, the measurement of shear statistics in redshift bins, or tomographic weak lensing, can greatly improve 
cosmological constraints \cite{Hu99}, especially for dark energy \cite{TJ04},
since the weak lensing signal scales differently with redshift
 for different models of dark energy evolution. 

The accuracy of cosmic shear constraints is affected by systematic effects. 
Systematics arise primarily from intrinsic alignments,
redshifts uncertainties, and shear measurement errors.
Intrinsic alignments are non-cosmological sources of shear correlation,
caused by an intrinsic correlation between the orientation of galaxies \cite{Crietal01}
or by shear-shape correlation, i.e., by a correlation between
galaxies shapes and surrounding density fields \cite{HiSe04}.
These effects have been modeled, eg., in Heymans et al. (2006b) and have also been constrained with data \cite{Maetal06}.
In particular, it has been shown \cite{Hietal07,BrKi07} that ignoring the effect of the  shear-shape
correlation on the amplitude of the two-point shear statistics heavily
biases the constraints on the equation of state of  dark energy.
The impact of redshifts uncertainties on cosmic shear has also been analysed
in several studies \cite{IH05,MHH06,vWetal06}

This paper deals with the third primary effect: shear measurement errors due to
the lack of an unbiased method to estimate
gravitational shear from astronomical images of galaxies. It aims to derive the effect of shear
measurement bias on dark energy constraints both with and without tomography.
The estimation of gravitational shear is technically
challenging and the methods available today do not seem to do better than
the percent level of accuracy. 
The Shear Testing Programme (STEP),  and more recently the gravitational
lensing estimation accuracy test (GREAT08) (Bridle et al. 2008), represent to
date the largest collaborations aimed at testing and improving
the accuracy of the existing shear measurement methods with the use of realistic sets
of simulated images. 
Current methods are based on the KSB approach
\cite{Kaetal95,LK97}, shapelet decomposition \cite{BeJa02,ReBa03,Ku06}, or Bayesian techniques \cite{Kietal08b}.
The STEP collaboration has
shown that the methods developed in the past years underestimate on average the shear
signal by a few percent \cite{Heetal06a,Maetal07a}, implying the
measurement of two-point shear statistics is underestimated by two to ten percent.
Moreover, the STEP collaboration showed that this bias is not constant but 
depends on the brightness and size of the galaxies. Indeed, for most methods the
measured shear is overestimated for bright galaxies and greatly underestimated
for faint galaxies. This may be a limitation of the accuracy with which one can
derive cosmological constraints using cosmic shear statistics in future ground-based
and space-based
astronomical surveys such as KIDS, LSST, Pan-STARRS, DES, JDEM or EUCLID.

There are in the literature various studies dealing with the impact of shear
measurement errors and PSF modeling on cosmic shear measurements
\cite{HiSe03,Ho04,vWetal05,JJB06,PHetal08}. There are also several cosmic shear forecasts 
of cosmological parameters which include and model generic types of systematic
 errors, such as multiplicative and
additive errors in shear measurements
\cite{Hutetal06,AmRe07,Kietal08a}. In these previous works  the bias
has been assumed to be a generic function of the redshift; in this paper we adopt a different approach, consisting in
explicitly quantifying the bias as a function of redshift using realistic
simulations, as opposed to using generic modeling. In order to derive the
expression of the bias as  a function of redshift we reanalyse the STEP2
simulations using a KSB pipeline. We observe that the bias depends on the
characteristics of the galaxies such as magnitude and size and we use these
dependencies  to derive a realistic expression of the  bias as a function of
redshift.
In order to evaluate the impact of this bias on the estimation of cosmological parameters
 we perform a likelihood analysis of tomographic two-point cosmic shear
in a space of cosmological and bias parameters, including small-scale
non-Gaussian corrections in the covariance matrix.
We note that even though we model the bias using a particular
implementation of KSB, the results we find are  general as 
all KSB methods tested by the STEP collaboration show the same
behavior.  

The paper is organized as follows: in Section \ref{step} we 
describe the results of the STEP2 analysis and the procedure to
estimate the bias as a function of redshift. Section \ref{theory} defines
the tomographic
two-point shear statistics in the presence of a redshift-dependent bias. In
Section \ref{constraints} we define the
 space of cosmological and shear measurement
bias parameters  where the likelihood analysis will be performed. Section \ref{covariance}
presents the correction for non-Gaussianity to be applied to the covariance
matrix of tomographic two-point correlation functions computed for a Gaussian
shear field. The derivation of this covariance in the real space is detailed in the Appendix.
The results of the likelihood analysis are presented and discussed in Section
\ref{results} and we conclude in Section \ref{conclusions}.

\section{STEP2 analysis}\label{step}

\begin{figure*}
\begin{tabular}{cc}
\psfig{figure=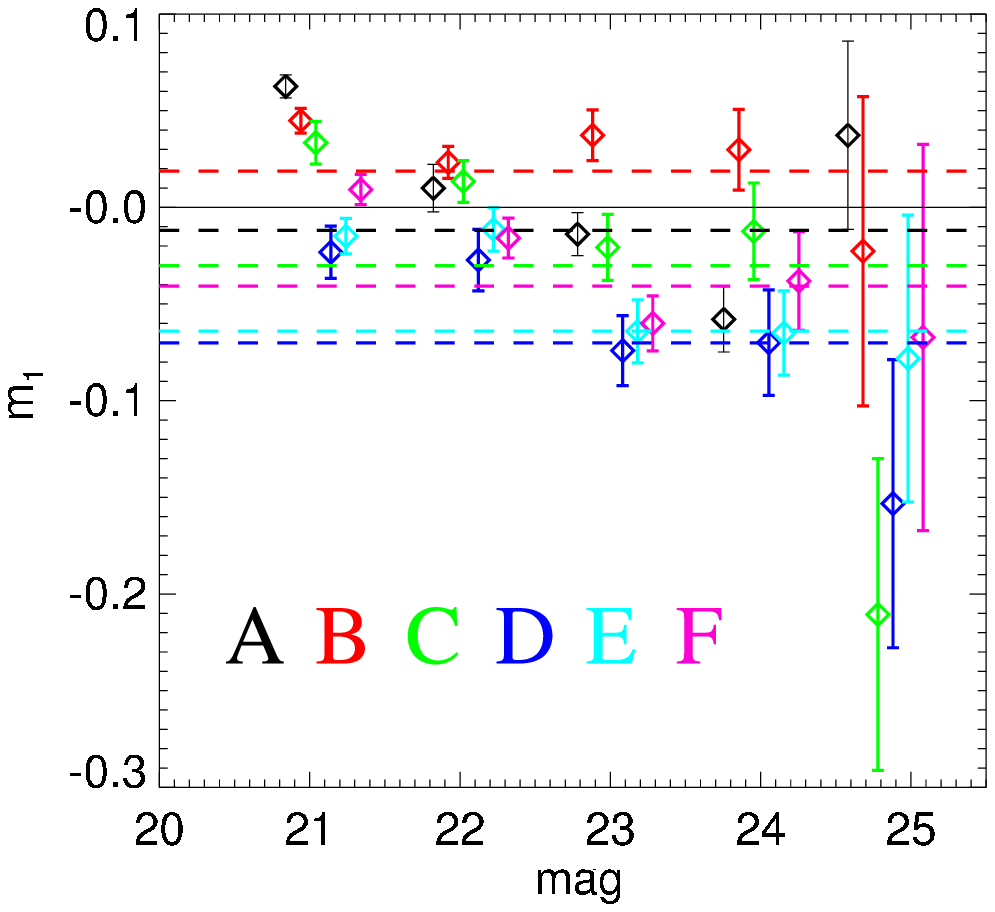,width=.45\textwidth}&\psfig{figure=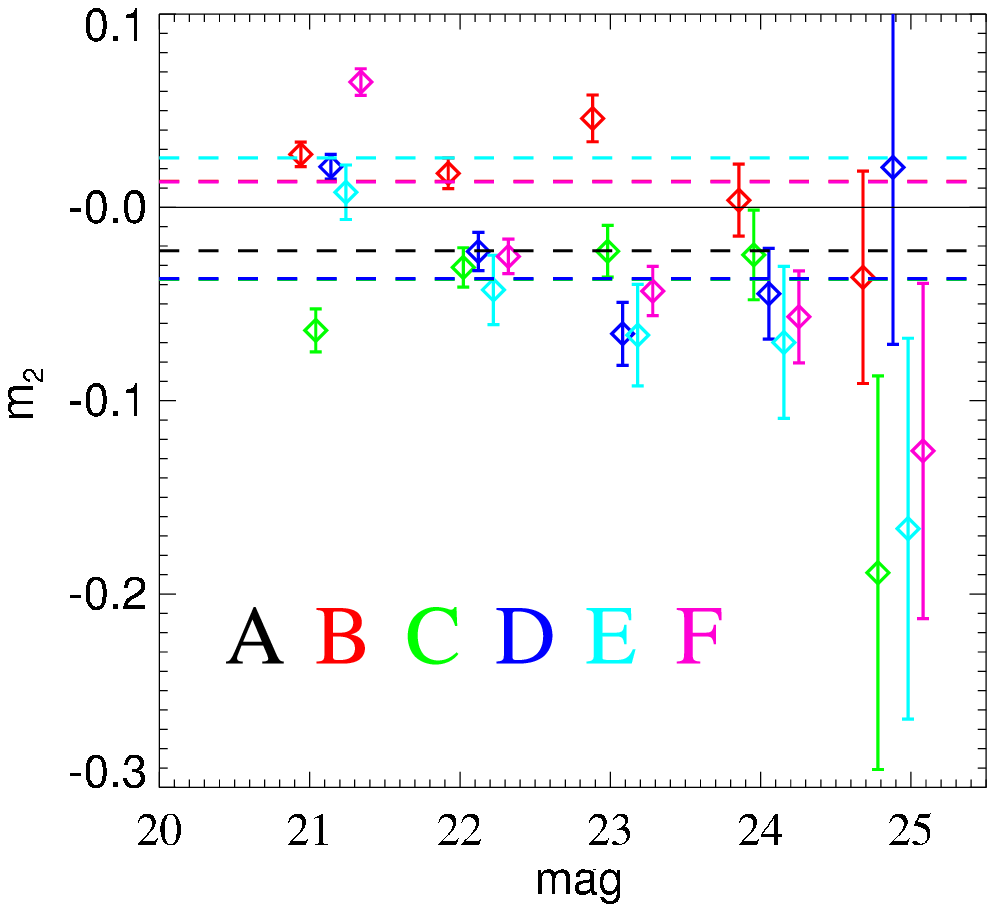,width=.45\textwidth}
\end{tabular}
\caption{\label{mag_bias} Left panel: values of the multiplicative bias $m_1$ as
  a function of magnitude  averaged on the 64 image pairs. Error bars represent the
  standard deviations  around the average value. Different colors refer to
  different PSF sets. The dashed lines show the  values of the multiplicative bias factor $m_1$ averaged on the whole catalogue. Right
  panel: same as  the left panel but for the second component $m_2$ .}
\end{figure*}

\begin{figure*}
\begin{tabular}{cc}
\psfig{figure=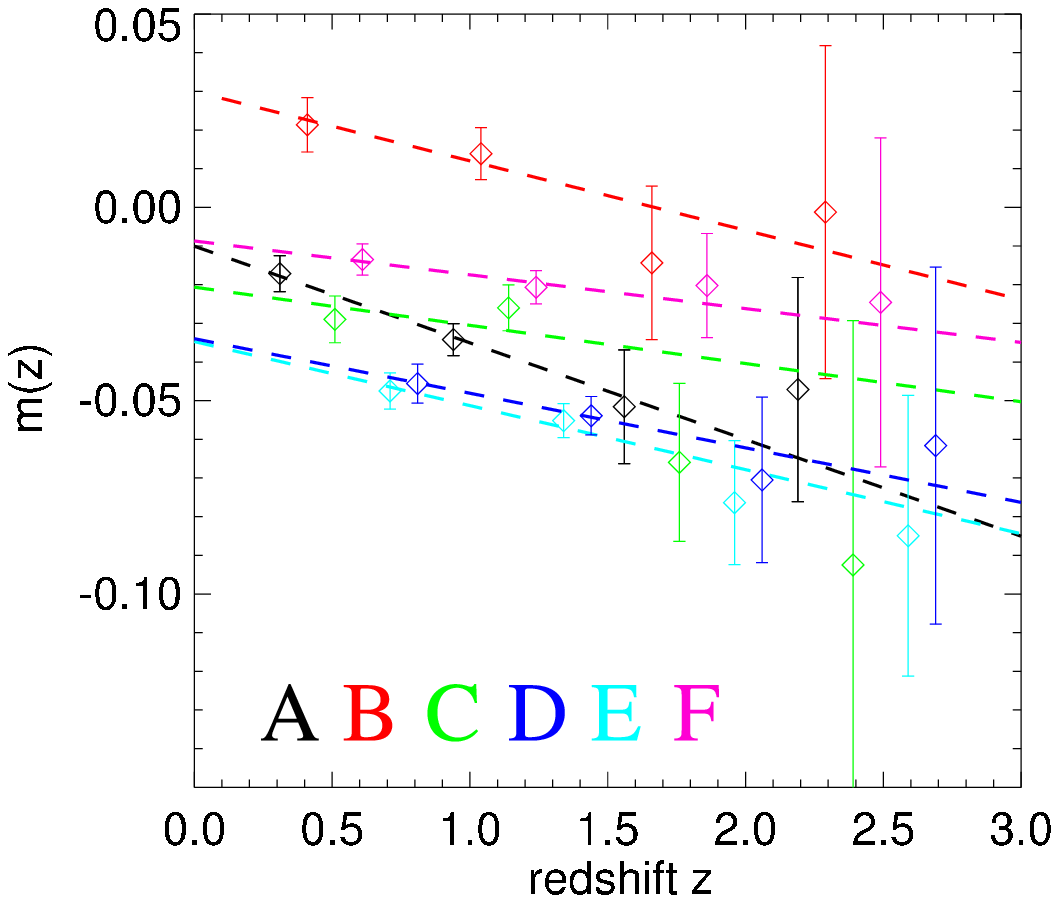,width=.45\textwidth}&\psfig{figure=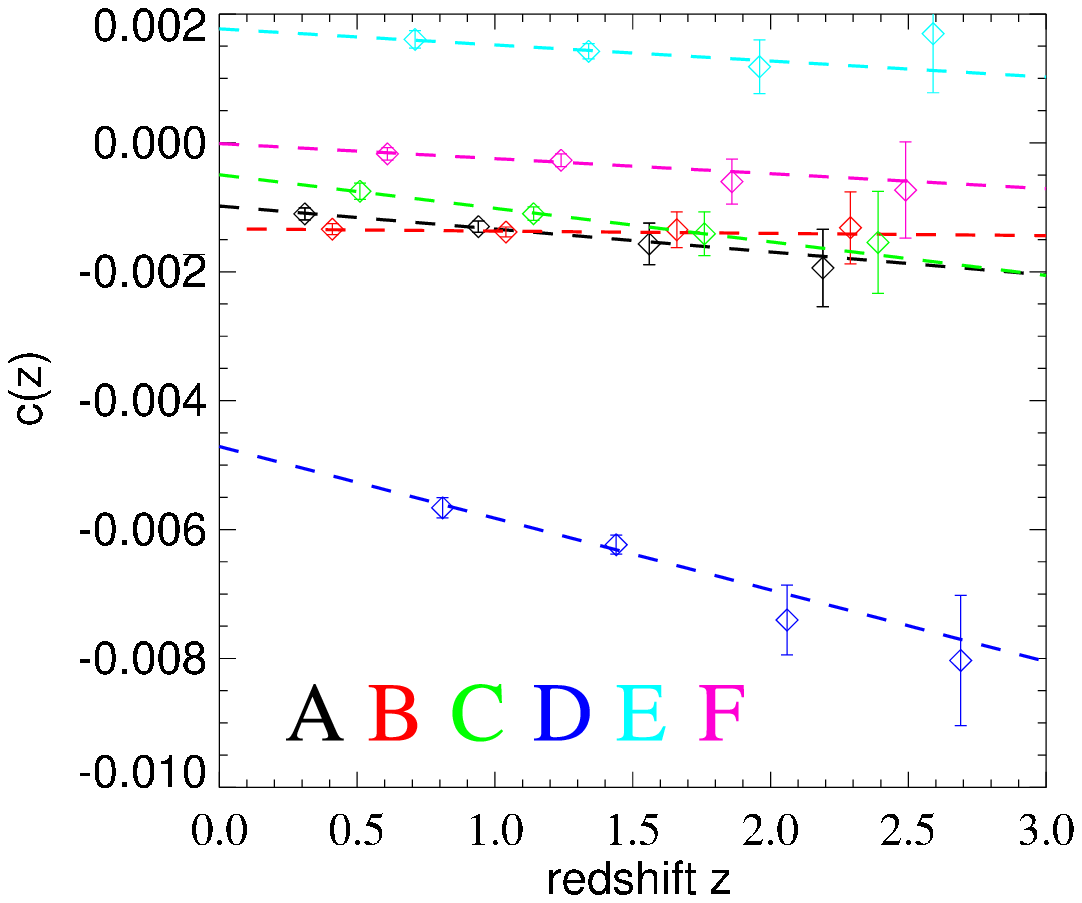,width=.45\textwidth}
\end{tabular}
\caption{\label{redshift_bias} Left panel: average value  of the multiplicative bias factor $m$ as a function of
  the redshift for each set of PSF. The  value has been obtained by averaging over the one hundred bootstrap realisations built as indicated in the text.
Error bars represent  standard deviations  of the one hundred realisations.  The dashed lines represent the best linear fit to the
  points. Right panel: same as the left panel for the additive bias constant $c$.}
\end{figure*}

In the last decade a large effort has been made in order to establish
the accuracy with which one can estimate shear from images of galaxies.
The Shear Testing Programme is a  collaboration which aims to test and
improve methods for PSF correction using realistic sets of simulations. 
The STEP1 and STEP2 simulations have a depth
similar to the images of the CFHTLS-Wide survey, which is to date  the largest
deep weak
lensing survey available. The results of the analyses on the simulations
can thus be immediately used to assess the accuracy on the cosmological constraints
from real surveys. The galaxies in the STEP2 set of
simulations have a brightness profile built using a shapelets decomposition of galaxies
observed with the Hubble Space Telescope \cite{Maetal04}; the resulting profiles are   
 more realistic than  the STEP1 simulations in which elliptical
 galaxies are characterised by de Vaucouleurs profiles and spiral
 galaxies are characterised by a bulge and a projected disk. 

We recall here the main characteristics of the STEP2 simulations
referring  the reader to Massey et al. (2007) for further information.
The STEP2 set of simulations is composed of six subsets, each characterised by a  different
 PSF.
Each subset is identified by an alphabetical letter from A to F. The PSFs A
 B and C are typical Subaru PSFs. The PSF D and E are highly elliptical, aligned along the
 $x$ and $y$ axis, respectively. The PSF F is a circularly symmetric
 PSF. Furthermore the PSF is constant across the field; the
 seeing size is $\sim 0.6$ arcsec, apart from PSF C for which the seeing is
 $\sim 0.8$ arcsec.
 For each PSF there are 64  pairs of $7^\prime \times 7^\prime$ images with pixel
 size of $0.206$ arcsec. Each  pair
 corresponds to one image and its rotation by 90 degrees. A shear field and a
 PSF are applied to the image pairs. The image
 pairs have been generated so that one can estimate the shear using a galaxy
 and its rotated pair, reducing the effect of  shape noise on the measured shear.
  The STEP2 simulations are therefore more suitable for our study than the
STEP1 simulations both because the galaxy profiles
are more realistic and the bias can be determined to a higher level of
 accuracy. For this reason we will consider the STEP2 set of simulations as
 our main dataset,  but  we will  briefly discuss the results we
 obtain using the STEP1 set of simulations at the end of this section. 

We analyse the STEP2 simulations using the KSB implementation used in Fu et
al. 2008 but with the following  significant changes: 
\begin{itemize}
\item we do not introduce a weighting scheme, i.e. all the objects have the
  same weight
\item  we use the trace of the shear polarizability tensor $P_g$ in the shear
estimator for each object
\item  we select objects only according to their
signal-to-noise ratio. The signal-to-noise is defined by the $\nu$
parameter  provided by the IMCAT \footnote{\url{http://www.ifa.hawaii.edu/~kaiser/}, developed
  by Nick Kaiser.} shape measurement software.
\end{itemize}
We introduce these modifications as  we want to keep the PSF correction method as basic as possible
avoiding arbitrary ``ad hoc'' choices which might change  the results
significantly.
For each PSF, we analyse the 64 pairs of images, merging the catalogues of the
image pairs so that the final catalogue is free of spurious detections.

As suggested in Massey et al. (2007) we describe the difference between 
the measured shear $\vc \gamma$ and the input shear signal $\vc \gamma_{ \rm true}$  
using the multiplicative  $\vc m=(m_1,m_2)$  and  additive $\vc c=(c_1,c_2)$ bias factors:
\begin{equation}\label{bias_param}
\gamma_i(\vc \theta)-\gamma_{i, {\rm true}}(\vc \theta)=m_i ~\gamma_{i,{\rm true}}(\vc
\theta)+c_i~    \qquad  i=1,2 ~.  
\end{equation}
The estimated shear at a position $\vc \theta$ is affected by a multiplicative
bias $\vc m$. The main reason for this calibration error is that KSB methods do not completely
correct for the isotropic kernel of the PSF, leaving thus a residual which is
seeing and magnitude dependent. Since the seeing in the STEP2 simulations is
constant across the image, $\vc m$ is independent of position.
The additive bias $\vc c$ comes from incomplete correction of the PSF
anisotropy. This originates from the approximative nature of the KSB method and also from
the residuals due to the modeling of the variation of the PSF across the
image, which depend on the distribution of the observed stars. 
The obtained $\vc m$ and $\vc c$ biases depend on the magnitude and on the
size of the galaxies.  We define the size of an object as the characteristic
size $r_h$ of the Gaussian profile which best fit the brightness profile  as
defined by the IMCAT software.
We verify that our results are similar to those  obtained by  STEP2
 by measuring  the values of $\vc m$ and $\vc c$ as a function of
magnitude  for the two ellipticity components (see Fig.~\ref{mag_bias}). 
For example, the value of the $\vc m$ components averaged over all galaxies lies between $-7\%$
and $+2\%$ depending on the PSF subset, with a mean over PSFs of $\sim -3\%$.

\begin{figure*}
\begin{tabular}{cc}
\psfig{figure=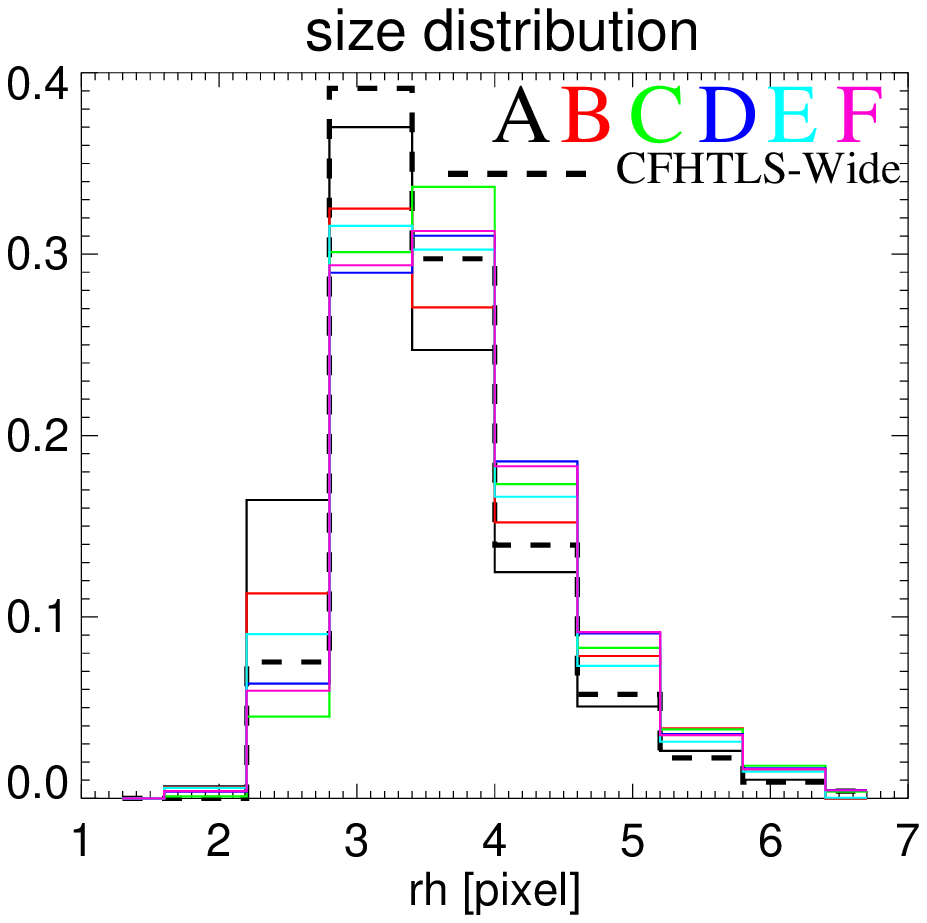,width=.45\textwidth}&\psfig{figure=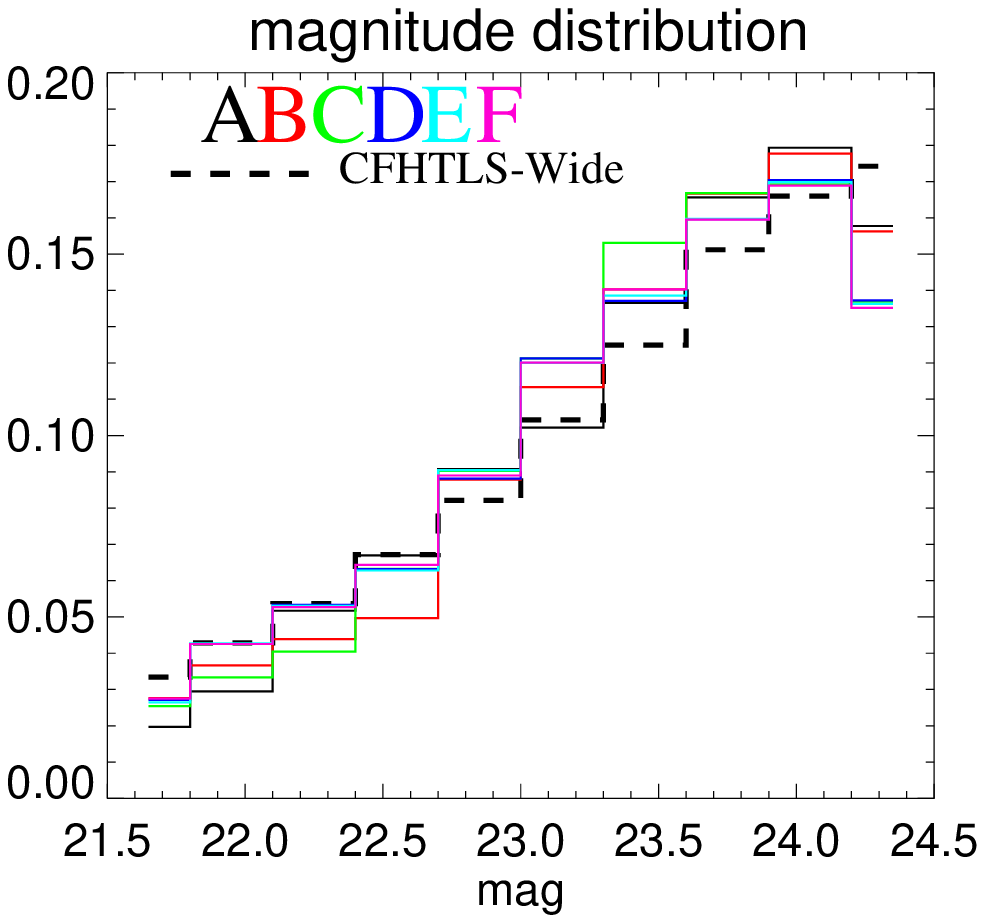,width=.45\textwidth}
\end{tabular}
\caption{\label{objects_distr}Left panel: size distribution of
  the galaxies in each of the STEP2 sets (solid histograms). The distribution of sizes
  of the CFHTLS-Wide galaxies is shown for comparison (dashed histogram). The
  size of the galaxies in the CFHTLS Deep catalogue have been converted
  into the STEP2 simulations pixel units of 0.206 arcsec. Right
  panel: distribution of magnitudes for each of the STEP2 PSF sets (solid)
  and for the CFHTLS-Wide galaxies (dashed). }
\end{figure*}
\begin{table}
\caption{\label{table1} Best-fit values of the redshift
  bias  parameters $a_m,b_m,a_c,b_c$ and their relative errors for
  each PSF set  of
  the STEP2 simulations. These values are used to plot the lines shown
  in  Figure~\ref{redshift_bias}. The last line shows the 
  averages over all PSFs.}
\begin{tabular}{l|c|c|c|c}
PSF& $a_{m}(10^{-2})$ & $b_{m}~(10^{-1})$ & $a_c ~(10^{-4})$ & $b_c ~(10^{-4})$\\
\hline
A   & $ -2.5 \pm 0.7$ & $-0.10 \pm 0.06$ &$-3.5 \pm 1.4$  & $-9.8 \pm 1.3$ \\
B &  $-1.8 \pm 1.1$& $0.28 \pm 0.09$& $-0.4 \pm 1.0$  & $-13.1 \pm 1.2$ \\
C & $-1.0 \pm 1.0 $& $-0.23 \pm 0.08$&  $-5.2 \pm 2.0$  & $-6.0 \pm 1.7$ \\
D  &  $-1.4 \pm 1.0$& $-0.41 \pm 0.07$& $ -11.1\pm 2.7$  & $-52.6 \pm 2.2$ \\
E & $-1.7 \pm 0.8$& $-0.41 \pm 0.06$&  $-2.5 \pm 2.3$  & $16.7 \pm 1.9$ \\
F & $-0.9 \pm 0.7$ & $ -0.11 \pm 0.06$& $-2.3 \pm 1.7$  & $-0.8 \pm 1.3$ \\
\hline
mean & $ -1.6 \pm 0.4$ & $-0.02 \pm 0.03$ & $-4.2 \pm 0.8$ & $-10.9 \pm 0.7$
\end{tabular}

\end{table}

We want now to convert the bias as a function of magnitude and size into a bias
as a function of redshift. 
To do so, we need to find the average redshift distribution of the galaxies
belonging to each size and magnitude bin, 
using   the CFHTLS-Deep photometric redshift catalogue \cite{Iletal06}.
For this we first convert the STEP2  magnitudes into the
corresponding CFHTLS-Deep values. This is done by simply adding the
photometric zero-point specified in the header of the images.
We then find the conversion between the  observed size of an object in
each set of simulations and the size it would have in the CFHTLS-Deep
catalogues. A conversion is needed because the pre-seeing size is the reference size which one
should use to compare populations of galaxies between catalogues with different seeing.
This is in principle  not straightforward  because the observed size depends on
the  seeing through a convolution.  
 We check the difference of the seeing values comparing the locus of the stars in
the simulations and in the CFHTLS-Deep catalogues. To do this we
 express the size of the stars in the CFHTLS Deep catalogue  in units of pixels
 of the  STEP2 simulations (the pixel size of the STEP2 simulations is 0.206
 arcsec whereas the pixel size of the  CFHTLS images is 0.186 arcsec).
  We find that the  difference between the star size in each set of
 simulations  and the star size in the CFHTLS-Deep catalogue is always
 smaller than $1/5$ pixels except for the PSF C where the difference is about
 half a pixel. From this,  we conclude that we do not need to reassign the size of the STEP2
 galaxies. 

We chose a magnitude range  $21.5 < {\rm mag} < 24.5$, for which a survey with
the same depth of the simulations is complete.
We divide the galaxies in bins of size and magnitude. 
 By the definition (Eq. \ref{bias_param}), $\vc m$ and $\vc c$ are
 two-dimensional vectors but for our purposes we  set  both components
 equal to the average  between the two  components  and define scalar
 quantities: $m=\frac{1}{2}(m_1+m_2)$ and $c=\frac{1}{2}(c_1+c_2)$. For each bin, $i$,
containing $n_i$ galaxies, we compute the average bias parameters 
$m_i$ and $c_i$  and respective uncertainties $\delta m_i$ and $\delta c_i$. We divide the CFHTLS-Deep catalogue into the same magnitude-size bins. 
Each galaxy in the simulation is assigned a value $m_j$ and $c_j$,
extracted from two Gaussians of mean  $m_i$ and $c_i$ and dispersion $\delta
m_i$ and $\delta c_i$, 
and a redshift $z_j$ extracted randomly from the CFHTLS-Deep galaxies
belonging to the same bin.
We bin the result in redshift bins $z_k$, and the total bias in each redshift bin $z_k$ is given by:
\begin{equation}\label{bin_bias}
m(z_k)=\frac{\sum_{j=1}^{n_k}m_j}{n_k},~~~~c(z_k)=\frac{\sum_{j=1}^{n_k}c_j}{n_k} 
\end{equation}
where $n_k$ is the  number of galaxies in the redshift bin.
We repeat this procedure one hundred times.
The final value of $m(z)$ and $c(z)$ and the relative errors are obtained
averaging over the one hundred realisations.  

We chose to fit $m(z)$ and $c(z)$ using a linear function:
\begin{equation}\label{linear}
m(z)=a_m~z+b_m,~~~~c(z)=a_c~z+b_c
\end{equation}
We perform the fit using only objects with redshifts $z<2.5$. We show the
values of the multiplicative and additive biases as a function of the redshift
in Figure~\ref{redshift_bias} where  the best linear fit to the
data is also shown. 
The best-fit parameter values and relative errors are summarized in Table
\ref{table1}.  Overall the shear is generally underestimated by a few percent.
 The multiplicative bias parameter  $m(z)$  has a negative slope $a_m$ which
 is roughly 1-2\% for all the PSF types.
 The additive constant $c(z)$ shows a slope which is consistent with zero
 except for PSF D. The value of the constant $b_c$ is also significant in
 particular for PSF D and E.
We repeated the analysis assuming the bias to depend only on the
magnitude, i.e., dividing the catalogue only in bins of magnitude,  and 
found  similar results for the functions $m(z)$ and $c(z)$. This is due to the
fact that the size distribution as a function of redshift does not change
significantly, so even though the bias slightly depends on the size the
average bias is roughly  the same for each redshift bin.

One should expect  the  values of the bias
parameters  to change if the selection criteria change. 
For example, including low signal-to-noise objects increases the overall bias 
and also affects the dependence on redshift, $a_m$.
In order to explore pessimistic cases we removed the
signal-to-noise threshold (spurious detections are still removed  
by merging each object with its rotated companion).  The biases
increase significantly. The mean $m$ over all PSFs is now larger than $10\%$,
whereas the redshift-dependence has a slope $a_m$ between $-6\%$ and $-4\%$, 
depending on the PSF. This is a consequence of the fact
that lowering the signal-to-noise threshold we include in the final catalogues
a larger number of faint objects which are preferentially at high
redshifts. This result shows that when using simulations to find a
recalibrating factor, capable of compensating an average shear measurement bias, 
the factor obtained can only be safely applied to a similar population of
galaxies. Likewise, our results are only strictly valid for surveys with
size and magnitude distributions similar to the STEP2 catalogue.
As we already mentioned, this is likely to be the case for the CFHTLS-Wide,
which has a similar depth to the STEP2 simulations.
We show in Figure \ref{objects_distr}
that the size and magnitude distributions of STEP2 indeed match those of the CFHTLS-Wide.

We  analysed the STEP1 sets of simulations and derived their
redshift-dependent biases.  We  notice that  the distribution of galaxies in
STEP1 includes a higher fraction of  large galaxies than real data
for the same depth such as the CFHTLS-Wide data.
 We find that the average multiplicative bias is  $\sim 2\%$, averaged over the
PSFs subsets with much less dispersion between PSFs than for STEP2. It is characterised by a  slope value $a_m$  always smaller than $1\%$, indicating that the accuracy of the shape
measurement degrades less rapidly when the signal-to-noise ratio decreases
in comparison to the STEP2 simulations.  

The analysis of STEP1 and STEP2  simulations give the same
qualitative results: the average bias is negative and the bias increases with
redshift. The amplitude of the  bias measured in the STEP1 simulations is  smaller than the
bias in the STEP2 simulations. The difference in the results is not surprising
as the shape and the amplitude of the measured bias depends on the
characteristics of the simulations. We conclude that in order to quantify the effect of the
bias on
real data one should use simulations that are as realistic as possible and as
we already pointed out the
STEP2 simulations are more realistic than the STEP1 simulations.

\section{Gravitational Lensing Statistics}\label{theory}
 \begin{figure*}
\begin{tabular}{cc}
\psfig{figure=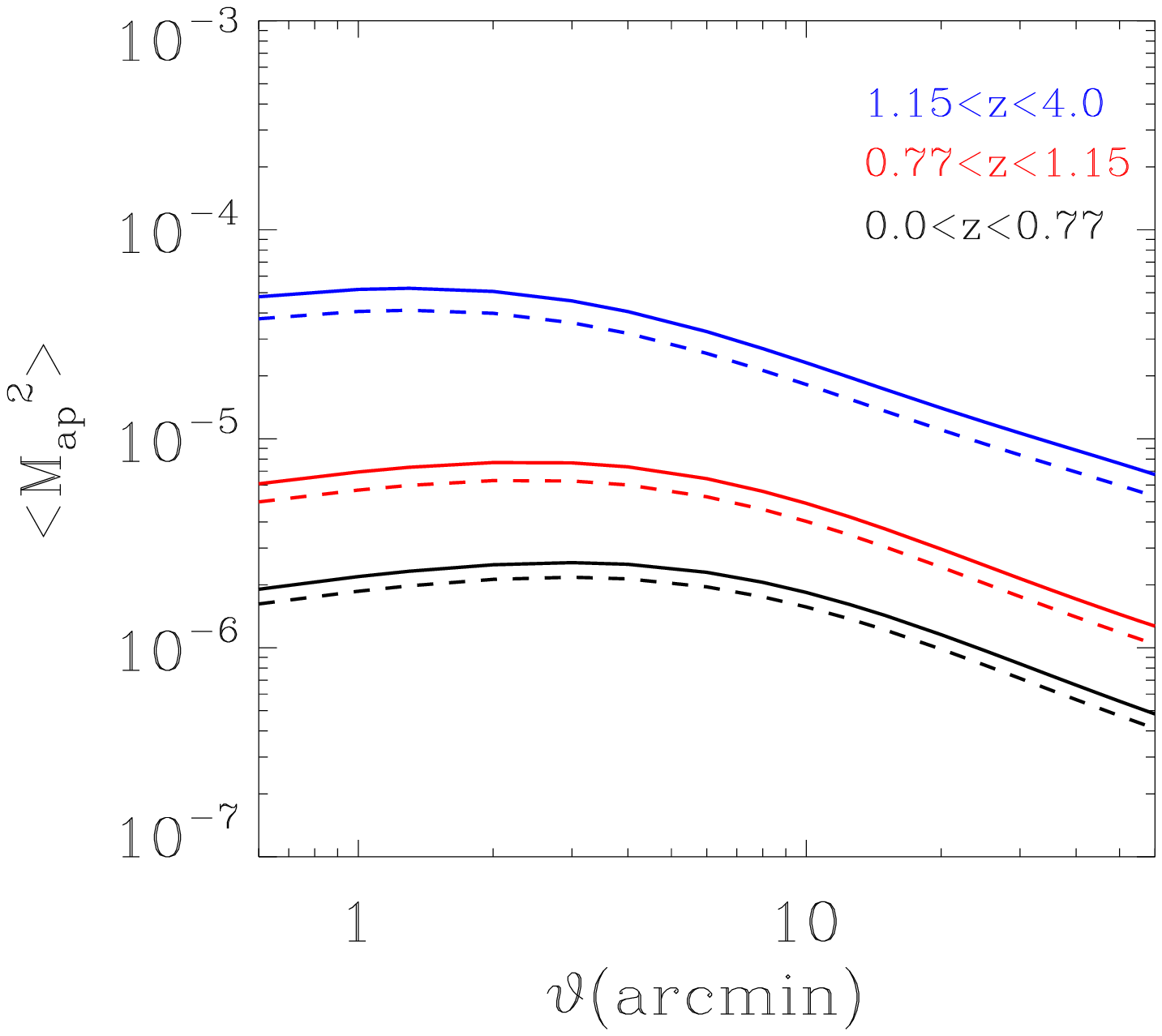,width=.45\textwidth}&\psfig{figure=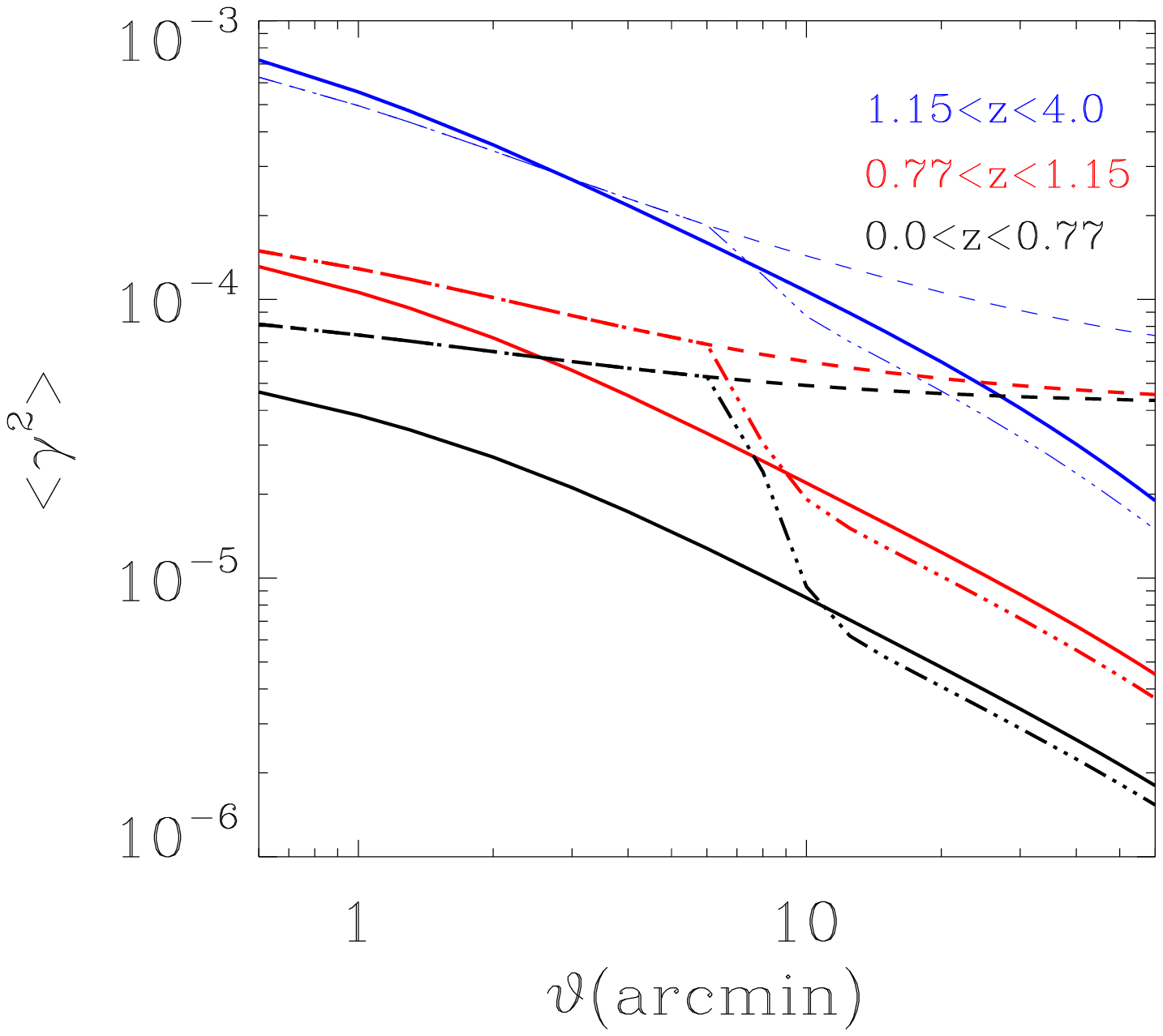,width=.45\textwidth}
\end{tabular}
\caption{\label{signal} Biased and unbiased two-point shear statistics as
  function of angular scale $\vartheta$ for three redshift bins. 
  The unbiased signal (solid lines) is compared with
  the signal that would be measured on a survey with very elliptical PSF (dashed
  lines) characterised by the following bias parameters: $a_m=-0.03$,
  $b_m=-0.06$, $ a_c=-0.001$, $b_c=-0.006$. Left panel: for $\langle \map
  (\vartheta) \rangle$. Right panel: for the top-hat  variance $\langle
  \gamma^2 (\vartheta) \rangle$ . The extra lines (dash-dotted) show a case
  where the additive bias only correlates on small scales. See text for details.}
\end{figure*}

In the next sections we explore the impact of the shear bias on cosmological
constraints obtained with two-point shear statistics, both with and without
considering redshift binning.  To
do so we derive an expression for two-point shear statistics which
include the redshift-dependent bias of the shape measurement.

From Eq.~(\ref{bias_param}), the measured correlation function is related to the true correlation
function via
\begin{align} \label{corr}
\langle \gamma\,(\theta)~\gamma^*(\theta') \rangle &= \langle
	[1+m(z)][1+m(z')]\gamma_{\rm true}\,(\theta)~\gamma^*_{\rm
	true}(\theta') \rangle + \nonumber \\
&+ \langle c\,(z)~c\,(z')\rangle\,. 
\end{align}
Since the two types of biases, $m$ and $c$, are
independent there are no mixed terms in Eq.~(\ref{corr}).
The first term of the right-hand side is identical to the correlation
function, $\langle \kappa_{\rm mod}(\theta)\,\kappa_{\rm mod}(\theta')\rangle$,
of a modified convergence $\kappa_{\rm mod}=(1+m)\kappa_{\rm true}$.
The power spectrum of the modified convergence relates to the matter power
spectrum. Using the Limber approximation \cite{BaSch01}, the relation is:

\begin{equation}\label{pkappa}
P_\kappa(s)=\frac{9}{4}\Om^2\Big(\frac{H_0}{c}\Big)^4 \int_0^{w(z_{\rm sup})} dw~
\frac{g_i(w)\,g_j(w)}{a^2(w)} P_\delta\Big( \frac{s}{f_K(w)};w \Big)~,
\end{equation}
with
\begin{equation}\label{model_red_new}
g_i(w)=\int_{w'}^{w'(z_{\rm sup,i})} dw^\prime ~[1+m(w^\prime(z))]~
p_{w,i}(w^\prime) \frac{f_K(w^\prime-w)}{f_K(w^\prime)}~.
\end{equation}
 $f_K(w)$ is the comoving angular-diameter
distance and $m(w)$ is the multiplicative bias written as a
function of the comoving distance $w$. Eqs.~(\ref{pkappa}) and
(\ref{model_red_new}) take into account tomography, describing
auto- and cross-correlations between redshift bins $[z_{\rm inf,i},z_{\rm
    sup,i}]$ where the distribution of the sources $p_{w,i}(w)$ is zero for $w<w(z_{\rm inf,i})$ and
$w>w(z_{\rm sup,i})$.
Notice the effect of the multiplicative redshift-dependent bias is
equivalent to a change in the source redshift distribution. Since $1+m(z)$ is
a decreasing function (cf. Fig.~\ref{redshift_bias}), the effective redshift
distribution is shallower.

The modified two-point shear measured in apertures can be computed as usual,
by integrating the shear-shear correlation function, Eq.~(\ref{bias_param}), with the appropriate
filters. For the top-hat variance at an angular scale $\vartheta$ we obtain 
\begin{align}
\langle   \gamma^2 ( \vartheta ) \rangle  &= 2 \pi \int_0^\infty ds s
P_\kappa (s)  [ W_1(s \vartheta) ]^2 + \nonumber \\
&+ 
\frac{1}{\pi^2 \vartheta^4} \int_0^\vartheta d^2 \theta
\int_0^\vartheta d^2 \theta'\left[
 \int_{w(z_{\rm inf})}^{w(z_{\rm sup})} dw~p_w(w) \right. \nonumber \\
& \left. \int_{w'(z'_{\rm inf})}^{w'(z'_{\rm sup})} dw'~p_w'(w')
 \,\langle \,c~(w(z)\,)\, c~(w'(z)\,)\,\rangle \right].
\label{model_top}
\end{align}
The first term integrates the modified convergence power spectrum,
Eq.~(\ref{pkappa}), using the top-hat filter in Fourier space,
$ W_1(\eta)=J_1(\eta)/(\pi \eta)$,
where $J_1(\eta)$ is the Bessel function of first order.
The second term of Eq.~(\ref{model_top}) is the contribution from the additive
bias. It is the integral in the aperture of the second term of
Eq.~(\ref{bias_param}), using the top-hat filter in the real space. Notice
that due to the redshift dependence of $c$ the integral over the aperture
also includes an integration in the radial direction, weighted by the source
distribution. If the correlation $\langle c\,(z)\, c\,(z')\,\rangle$ is constant
across the image, the effect of
the additive redshift-dependent bias is to add a constant to
the two-point shear statistics.

Similarly, the variance of the aperture mass is
\begin{equation}\label{model_map}
\langle   \map ( \vartheta ) \rangle  = 2 \pi \int_0^\infty ds s
P_\kappa (s)  [ W_2(s \vartheta) ]^2~,
\end{equation}
with $ W_2(\eta)=12\,J_4(\eta)/(\pi \eta^2)$, where $J_4(\eta)$ is the Bessel function of fourth order.
In this case there is no contribution from the additive bias because the
filter of the aperture mass dispersion in real space is a compensated
filter and thus insensitive to a constant shear.

Figure~\ref{signal} shows biased and unbiased two-point shear
statistics for three redshift bins. Only the three auto-correlations are
shown.  For this example we plot  Eqs.~(\ref{model_top}) and (\ref{model_map})
using a cosmological fiducial model, and 
bias parameters similar to PSF D and E, which are the most extreme cases. 
We  parameterise the  distribution of the sources, $p_w(w)$, as in Benjamin et
al. (2007) 
\begin{equation}\label{redshift}
p(z)=\frac{1}{N}\frac{z^\alpha}{z^\beta+z_0}
\end{equation} 
normalized to 1, 
with  $\alpha=0.73,~\beta=4.52,~z_0=0.80$, corresponding to a median
redshift $z_m=0.77$. These values were derived by fitting the density
of galaxies $n_k$  found in Section \ref{step} for each PSF, using only
galaxies with $z<2.5$. We find similar results for all PSFs. 

The left panel of  Fig.~\ref{signal} shows that the modified aperture mass
variance decreases by a scale independent factor of $15\%$ to $30\%$, 
 increasing with redshift. This increase produces a shallower scaling of the amplitude of  $\langle
\map(\theta,z)\rangle$ with the redshift. Modeling the redshift scaling
with a power law we find it changes as:
\begin{eqnarray}
&&\langle   \map ( \vartheta ) \rangle \propto \bar z^{2.05}\qquad\qquad{\rm (no~ bias)} \nonumber\\
&&\langle   \map ( \vartheta ) \rangle \propto \bar z^{1.95}~~~ ( a_m=-0.03,~b_m=-0.06).\nonumber
\end{eqnarray}

The right panel of Fig.~\ref{signal} shows that the top-hat variance is dominated by the additive
constant. Indeed, assuming the additive bias is constant across the image and correlates at all scales, the
second term of the right-hand side of Eq.~(\ref{model_top}) writes 
\begin{equation}
\mathcal C = \left [\int_{w(z_{\rm inf})}^{w(z_{\rm sup})} dw~ c~(w(z))~
  p_w(w)\right ]^2~.
\end{equation}
A constant bias of $\abs{b_c}\approx 10^{-3}$, which is the average value between the
PSFs shown in the Table \ref{table1},  yields ${\mathcal C} \sim  10^{-6}$
which is significant when compared to the cosmic shear signal, especially at the lower
redshift bins, invalidating any cosmological interpretation. This is in
agreement with Huterer et al. (2006) where it was found that an additive bias
larger than $\sim 10^{-4}$ would degrade the cosmological constraints of
future weak lensing probes.
This result suggests that whenever the PSF is not well corrected and the
residual is  constant  aperture mass statistics should be preferred.

We show in the right panel of  Figure~\ref{signal} a more realistic case of modified
top-hat variance, which assumes the PSF is modeled independently in sub-regions
of the image. In this case, the additive bias only correlates at the scales of
those sub-regions. In this example we assume the correlation is constant below $\sim 5\arcmin$, which is a
typical scale of the CCDs of ground-based cameras, and decreases steeply
(exponentially) to zero at larger scales. In this  toy-model  the PSF
residuals  are
strongly correlated at small scales enhancing strongly the signal, whereas on large scales $\vartheta >
5\arcmin$  the PSF residuals are decorrelated and only the
multiplicative constant affects the signal.  We expect something similar to
happen on real data as  the PSF varies 
across the field and it is generally corrected by interpolating the PSF shape
over a  characteristic area.  If this is the case, the additive constant would
affect both the top-hat and the aperture mass variances and the effect could
be significant  at small scales. However, the STEP2 simulations have a PSF which is constant across the field of view
and this does not allow one to explore in more detail the effect of such a
residual on real data.

\section{Parameter space definition}\label{constraints}
In order to estimate the impact of the redshift-dependent shear
measurement bias on cosmological
constraints, we perform a likelihood
analysis in  a nine-dimensional  grid using two-point shear statistics with
and without tomography.

  We define the log-likelihood of a model $\phi$ as 
\begin{align}\label{like} 
{\mathcal L}(\phi)
&=\frac{1}{2}\left[\boldsymbol{d}_{ij}(\vt)-\boldsymbol{v}_{ij}(\vt;\phi)\right]^t{\bf
  C}(\vt,\vt';\phi)^{-1} \nonumber \\
&\times
\left[\boldsymbol{d}_{ij}(\vt')-\boldsymbol{v}_{ij}(\vt';\phi)\right].
\end{align}
Both the tomographic vector of data, $\boldsymbol{d}_{ij}(\vt)$, and the
two-point functions theoretical predictions,  $\boldsymbol{v}_{ij}(\vt;\phi)$,
are computed at twenty angular scales, between 6 \arcsec and 1 \degr, and correlate three redshift bins $\,i,j=[1,3]$. 
The covariance matrix of the estimator used to measure the two-point shear
statistics, ${\bf C}(\vt,\vt';\phi)$, has dimension $120 \times 120$.

 The redshift bins are the ones used in Fig.~\ref{signal}:
$z_1=[0,0.77],~z_2=[0.77,1.15],~z_3=[1.15,4.0]$, with mean redshifts of $\bar
 z_1=0.47,~\bar z_2=0.94, \bar z_3=1.69$, respectively.
They were chosen such as to optimise the information. 
Since the signal increases with redshift, it is more efficient to
divide the redshift range such that the lower-redshift bins contain a larger
fraction of the galaxies. We followed the strategy of Hu (1999) and defined
the bins such that $z_1$ contains half of the galaxies and $z_2$ and $z_3$
contain one quarter each.  
The correlation coefficient between the aperture mass variances of the various
bins is defined as
$r_{ij}(\vt)={M_{\mathrm{ap},ij}^2}(\vt)/({M_{\mathrm{ap},ii}^2}(\vt)\,{M_{\mathrm{ap},jj}^2}(\vt))^{1/2}$. 
We measure a large correlation $r_{12}\sim r_{23} \sim 0.9$ between consecutive
bins, which shows that further subdivisions do not increase the amount of information.

The data vector, $\boldsymbol{d}_{ij}(\vt)$, is computed  for a fiducial
 model similar to the  WMAP5 average values ($\Omega_m=0.27$,~
 $\Omega_\lambda=0.73$,~ $h=0.705$,~$\sigma_8=0.812$,~$\Omega_b=0$) for a $\Lambda$CDM cosmology
\cite{Duetal08} and no bias. 

The value of $\boldsymbol{v}_{ij}(\phi)$ depends on cosmological parameters
and also on the bias parameters $a_m,b_m,a_c,b_c$.
The chosen range of the bias parameters covers the results of the STEP2 analysis: $a_m=[-0.03,0]$, $b_m=[-0.06,0.06]$, $a_c=[-0.001,0]$ and $b_c=[-0.006,0.002]$.
The  cosmological parameters varied in the grid are: the matter density
$\Om=[0.1,1.0]$,  the Hubble constant $h=[0.6,0.8]$,  the normalisation
of the matter power spectrum $\sigma_8=[0.5,1.1]$, and two
parameters, $w_0$ and $w_1$, for the equation of state of dark energy,
$w_{\rm DE}(z)=\rho_{\rm DE}(z)/P_{\rm DE}(z)$. We model $w_{\rm DE}(z)$ in
two different ways. The first one is a constant equation of state, $w_{\rm
  DE}(z)=w_0={\rm  const}$; for this model we vary $w_0=[-2.0,0]$.
The second model has been suggested by Benabed \& van Waerbeke (2004):
\begin{equation} \label{sugra} 
w_{\rm DE}(z)=\left\{
\begin{array}{l}  
w_0+w_1 \log(1+z)\,,\,z \le 1  \\
w_0+w_1[\log(2)-\arctan(1)+\arctan(z)]\,,\,z > 1
\end{array}\right.
\end{equation}
 Benabed \& van Waerbeke (2004) have shown this model is able to mimic quite 
 accurately the Universe dynamics generated by SUGRA 
 potentials. For this case we consider the ranges $w_0=[-2.0,0]$, $w_1=[0,0.4]$.

For the source redshift distribution we use the parameterization of
Eq.~(\ref{redshift}). We do not vary its parameters in the likelihood
analysis. This choice is supported by the fact that the errors affecting the estimation
of the parameters $\alpha$, $\beta$ and $z_0$  are about one percent, 
including the sample variance affecting the CFHTLS-Deep catalogue
\cite{vWetal06}. 
\begin{figure*}
\begin{tabular}{cc}
\psfig{figure=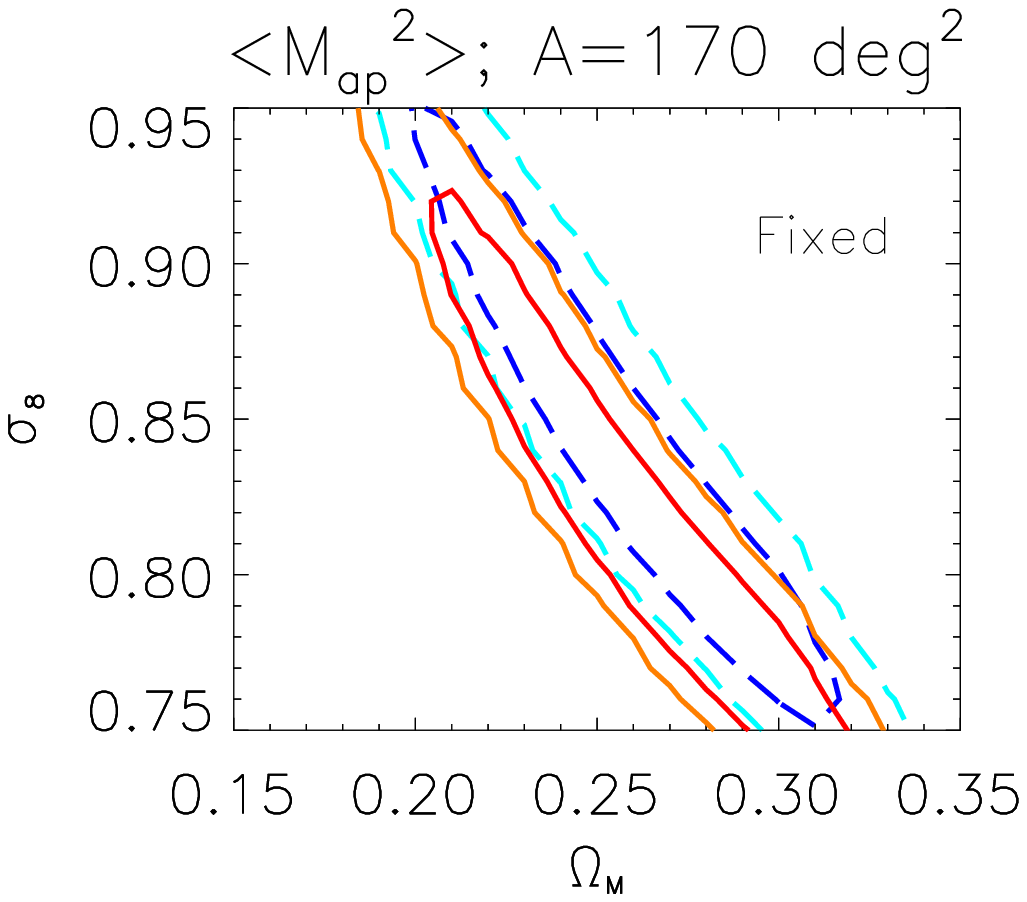,width=.45\textwidth}&\psfig{figure=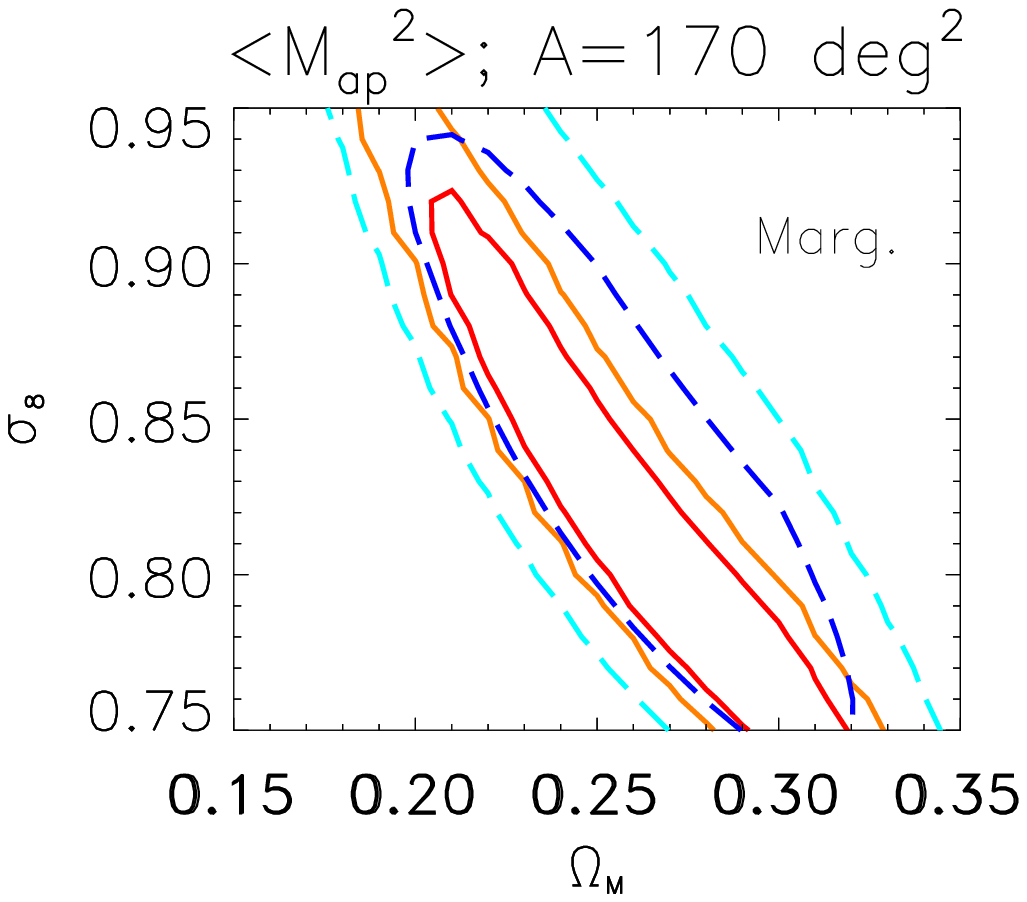,width=.45\textwidth}
\end{tabular}
\caption{\label{res_wide} Cosmological constraints in the $[\Om,\sigma_8]$
  plane  for a survey of area $170 ~{\rm deg^2}$ and three tomographic bins. The
  contours represent the 68\% and 95\% confidence regions and are marginalized
  over the hidden cosmological parameters. Left panel: the effect of a fixed bias of
  $a_m=-0.02,~ b_m=0$ (dashed, blue) is compared with the bias-free contour
  (solid, red). Right panel: the biased contour is now marginalized over the bias parameters
  (dashed, blue) and compared with the bias-free contour (solid, red).}
\end{figure*}
\section{Covariance Matrix}\label{covariance}
In this section we compute the covariance matrix ${\bf C}(\vt,\vt';\phi)$ for Eq.~(\ref{like}).
Most studies of weak lensing tomography have been made in Fourier
space where the covariance matrix is easy to compute. Analyses made in real space usually evaluate the
covariance matrix from Monte Carlo simulations (eg., Simon, King \& Schneider
2004), or more recently analytically from the covariances of
power spectra \cite{JSE08}.
Here we follow a different approach deriving analytically in  real space $C_{++}=C(\xi_{+,ij}(\vartheta_1),\xi_{+,kl}(\vartheta_2))$, 
the covariance matrix of $\xi_{+,ij}(\vartheta)$, extending to tomography the formulae derived
 in Schneider et al. (2002).  The details of the derivation are given in appendix A.  
The covariances of the estimators of $\langle \map ( \vartheta
 ) \rangle$  and $\langle   \gamma^2 ( \vartheta ) \rangle$ are afterwards derived from the
 covariance of $\xi_+(\vartheta)$ (Schneider et al. 2002).

The pure cosmic variance term is analytically derived in Appendix A assuming
 a Gaussian shear field
 [Eqs.~(\ref{cosmic_var_term1}) and (\ref{cosmic_var_term2})], for the a
 fiducial model defined in this case defined in the previous section.
We then
 recalibrate this matrix for non-Gaussian contributions following Semboloni et
 al. (2007), hereafter S07. This consists in multiplying the Gaussian cosmic
 variance by a calibration factor $\mathcal F(\vt_1,ij;\vt_2,kl)$. 
 
We found that the calibration factor found in S07 needed to be modified for
the case of tomography by applying it to Eqs.~(\ref{cosmic_var_term1}) and (\ref{cosmic_var_term2}),
and comparing the result with the non-Gaussian cosmic
variance for tomography, measured directly on the same set of ray-tracing simulations used in
S07. The best agreement is obtained when
the redshift used in the calibration factor of S07,
\begin{equation}\label{eq:fit_nogauss}
{\mathcal F}(\vartheta_1,ij;\vartheta_2,kl)=\frac{\alpha\,(z)}
{(\vartheta_1 \vartheta_2)^{\beta\,(z)}}\,,
\end{equation}
is redefined as an effective redshift $z=z(ij;kl)$ given by
\begin{equation}
z(ij;kl)\equiv z=\frac{1}{2}\sqrt{(\bar z_i+\bar z_j)\,(\bar z_k+\bar z_l)}\;.
\end{equation}
The effective redshift is thus the geometric mean of the average redshifts of
the bins involved in each of the two correlations $\xi_{+,ij}$ and $\xi_{+,kl}$. 
The functional form of $\alpha(z)$ and $\beta(z)$ are defined in S07, with
 the same best-fit parameter values $(a_1,a_2,a_3;b_1,b_2,b_3,b_4;t_1,t_2,t_3)$ indicated there.
Finally, the recalibration is only applied at angular scales below the $\vartheta_{\rm max}$ defined in S07.

We find that this fitting formula agrees with the measured covariance to an accuracy of $20\%$; the worst
result being for the lower redshift bin. This recalibration might
overestimate the values of the covariance matrix \cite{TJ08} (as the
simulations we used have a high value   $\sigma_8=1$) but  this is not so important as  we include in our likelihood   analysis large scales measurements.  

After recalibrating the cosmic variance  term, we add the other contributions to
 $C_{++}$: the coupling
between shape-noise and the shear signal given by Eq.~(\ref{mixed}), and the pure
shape-noise term given by Eq.~(\ref{noise}). These terms depend on 
the  density  of galaxies,  the total area of the survey and  the intrinsic ellipticity dispersion. We assume a
density  of galaxies integrated along the line of sight of $n=12/ {\rm
  arcmin}^2$ which is similar to the CFHTLS-Wide survey with a median
 redshift of $z_m=0.8$. To find the density of galaxies for each redshift bin, we normalise 
 the redshift distribution of the sources, $p(z)$, to $n$ and integrate it
 between each $z_{\rm inf}$ and $z_{\rm sup}$. We use an intrinsic ellipticity dispersion $\sigma_\epsilon=0.44$. 
The cosmological constraints are derived for two different survey areas:  $A=170 ~
{\rm deg}^2$  and $A=2000~ {\rm deg}^2$, which are a good representation of current and future weak lensing
 ground-based multicolor surveys. The first case has the same sky
 coverage and depth as the CFHTLS-Wide survey. The second has similar sky coverage to the upcoming KIDS survey, while being
 about half a magnitude deeper than KIDS. However, as the shape measurement
 bias is a function of signal-to-noise, we expect it to be
 larger, for the same magnitude, in a shallower survey than in a deeper one.

Finally, using Eq.~(42)  of Schneider et al. (2002) with $K_+=1$,  we 
 obtain the covariance matrix of the estimator of  $\langle   M_{\rm ap}^2 (
 \vartheta) \rangle$ from $C_{++}$, and using a similar expression
 we derive the covariance matrix of the estimator of $\langle \gamma^2
 (\vartheta) \rangle$. 

During the likelihood analysis the covariance matrix is
 kept constant, neglecting its dependence on the cosmological parameters.
The accuracy of this approximation is studied in Eifler, Schneider \& Hartlap (2008).
We also neglect its dependence on bias parameters. Thus in the  likelihood
 defined by  Eq.~(\ref{like}),  the dependence on the bias is only contained
 in ${v}_{ij}$. This type of   systematic error which  affects the
 signal but does not introduce extra noise in the covariance matrix  has been
 called Type I  by  Kitching, Taylor \& Heavens 2008a and it allows for self-calibration.

\section{Results}\label{results}

In this section we obtain cosmological constraints by performing the likelihood
analyses previously described, both with and without tomography. 
As stated earlier on, our approach is to
include the realistic bias measured in STEP2, with as little extra
modeling as possible. For this reason we choose not to model the correlation
of the additive term, which does not come directly from the STEP2 analysis. 
We will focus thus on the multiplicative bias and show constraints using the
aperture-mass dispersion only.

 \begin{figure}
\psfig{figure=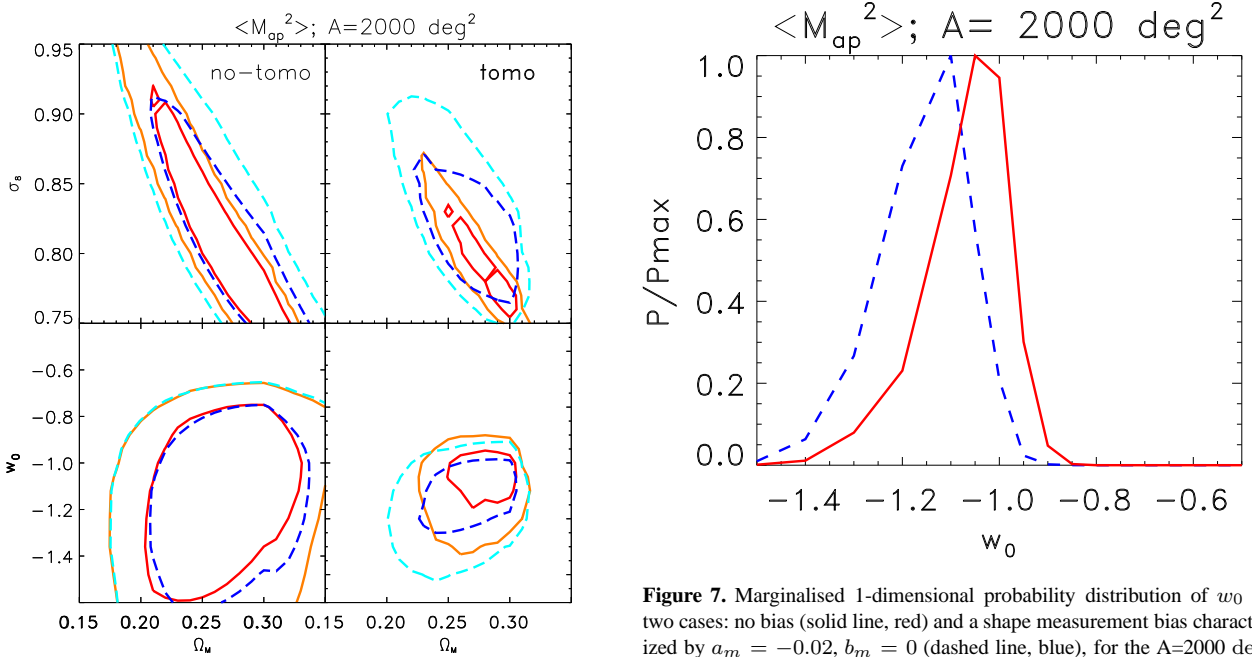,width=.50\textwidth}
\caption{\label{const_res} Confidence regions  ($68\%$ and $95\%$)  on the
  $[\Om,w_0]$ and $[\Om,\sigma_8]$ planes for the $2000~{\rm deg}^2$ survey. 
  The contours, marginalised over the hidden parameters, are shown for the
  case of no bias (solid, red) and marginalising 
  over the shape measurement bias parameters (dashed, blue).  Left panels show
  results of the non-tomographic analysis and right panels refer to the tomographic case. }
\end{figure}

We start by considering the $170~ {\rm deg}^2$ survey, which has relatively
large noise, to asses  if a redshift-dependent shape measurement
bias already affects the results of a tomographic analysis of a survey such as
the CFHTLS-Wide.
The constraints in the $[\Om,\sigma_8]$ plane, using auto- and
cross-correlation in the three redshift bins previously described,
are shown in  Figure~\ref{res_wide}. In the left panel, contours are shown for
the case of no bias and a particular choice of bias $(a_m=-0.02\,,\,b_m=0)$. 
The biased contour is clearly shifted
to the upper-right corner to compensate the negative multiplicative bias. 
This shows that our choice of redshift binning is quite effective, having a
noise level low enough for systematics to become important. 
The shift in the constraints can be quantified by fitting the direction of the
degeneracy seen in the plot. In the case of no 
bias the fit is given by
\be
\sigma_8(\Om/0.275)^{0.48}=0.79 \pm 0.02\,,
\ee
whereas in the case of this particular bias we find
\be
\sigma_8(\Om/0.275)^{0.43}=0.81 \pm 0.02\,.
\ee
This result shows that ignoring the presence of the shape
measurement bias would lead to biased contours.

\begin{figure}
\psfig{figure=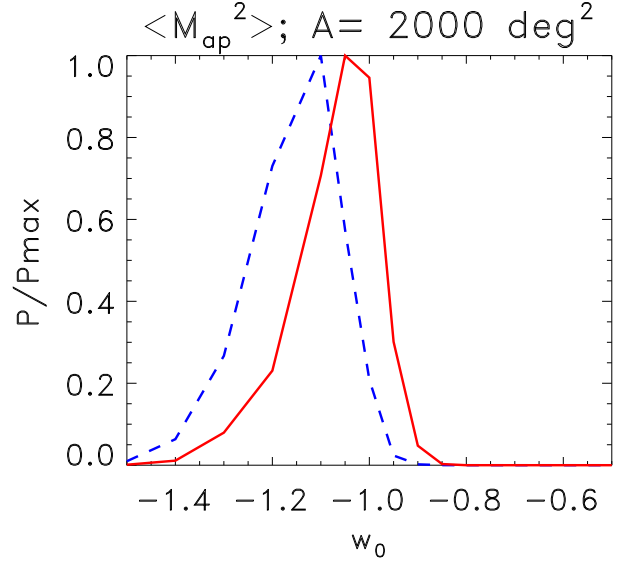,width=.45\textwidth}
\caption{\label{bias_w0} 
Marginalised 1-dimensional probability distribution of $w_0$ for two
cases: no bias (solid line, red) and a shape measurement bias characterized by $a_m=-0.02$, $b_m=0$
  (dashed line, blue), for the A=2000 ${\rm deg^2}$ survey. 
The peak of the probability distribution changes significantly
 with the shape measurement bias parameters.}
\end{figure}

When the amplitude of the bias is not known, the safer approach is to
marginalize over a reasonable interval of bias. Marginalizing over the range
suggested by our STEP2 analysis, we find the contour shown in
Figure~\ref{res_wide} (right panel). The precision of the tomographic measure
of $[\Om,\sigma_8]$ is reduced by a factor of 2.
This result includes the degeneracies between cosmological and bias
parameters. Marginalizing over the cosmological parameters, the bias parameters
could be estimated from the data, i.e., the signal could be self-calibrated. For
the self-calibration to be effective, priors would be needed. They can be
obtained with extra
information from independent measurements affected by the same systematics. Other
measures of cosmic shear such as higher-order correlation functions
\cite{BvWM97} or the shear-ratio test \cite{JT03} could be helpful for this purpose.

Finally, we also produced contours in the  $[w_0,\Om]$ plane.
In this case the systematics do not have a relevant effect, since the
bias-free contours are too broad and the estimate is noise-dominated.
 
We turn now to the case of the survey of $A=2000 ~ {\rm deg^2}$,  in
order to see how the shape measurement bias can affect the weak lensing
analysis of data from future ground-based observations. 
We quantify the constraints using a figure-of-merit, FoM,  defined as the inverse of
the area enclosed by the $68\%$ confidence level.
Quoted results imply a marginalisation of the bias over its full range.

In Figure~\ref{const_res}, 
we show the  cosmological constraints in the $[\Om,\sigma_8]$ and 
$[\Om,w_0]$ planes, both for tomography and non-tomography, using the variance
of the aperture mass.
The first interesting result is that tomography produces an enormous gain in
 all constraints, but especially in dark energy. Looking at the
bias-free contours, the increase in the FoM between using and not using
tomography is a factor of 10 for the $[\Om,w_0]$ contour and a factor of 6 for
 $[\Om,\sigma_8]$. This is larger than the improvement between the two surveys
 due to the increase in area, which scales with the square root of the
 survey area and is thus around 3.5.

 \begin{figure*}
\begin{tabular}{cc}
\psfig{figure=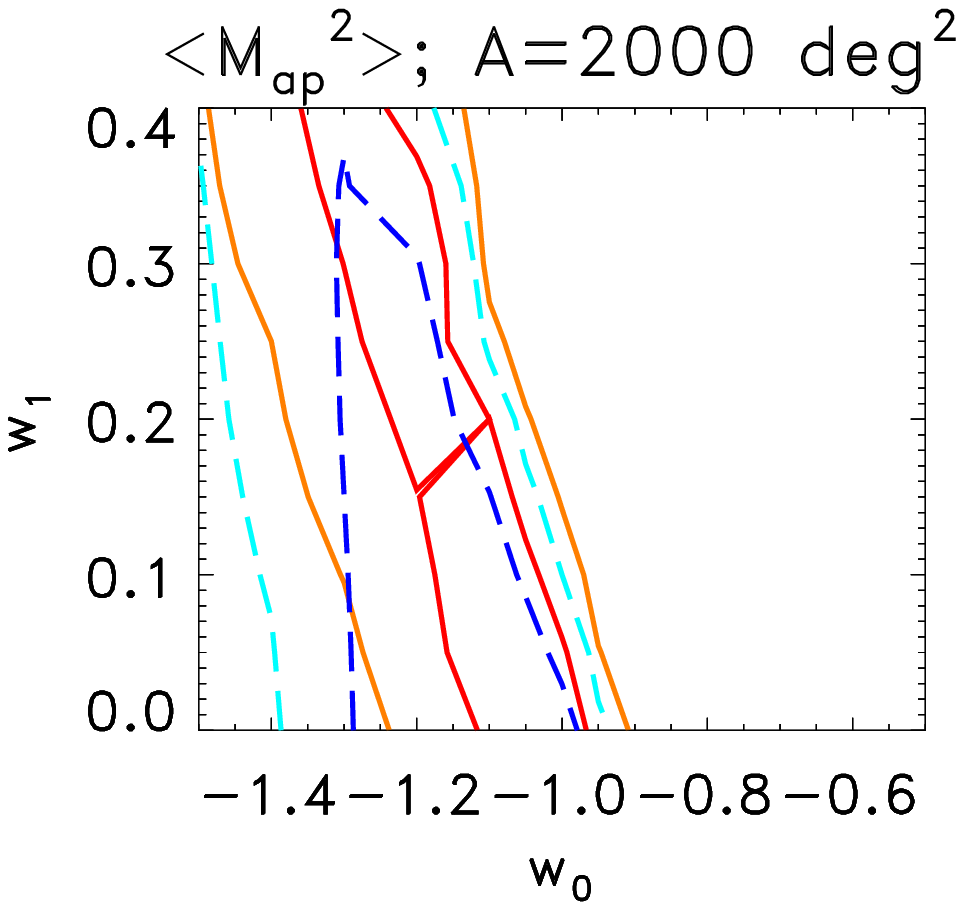,width=.45\textwidth}&\psfig{figure=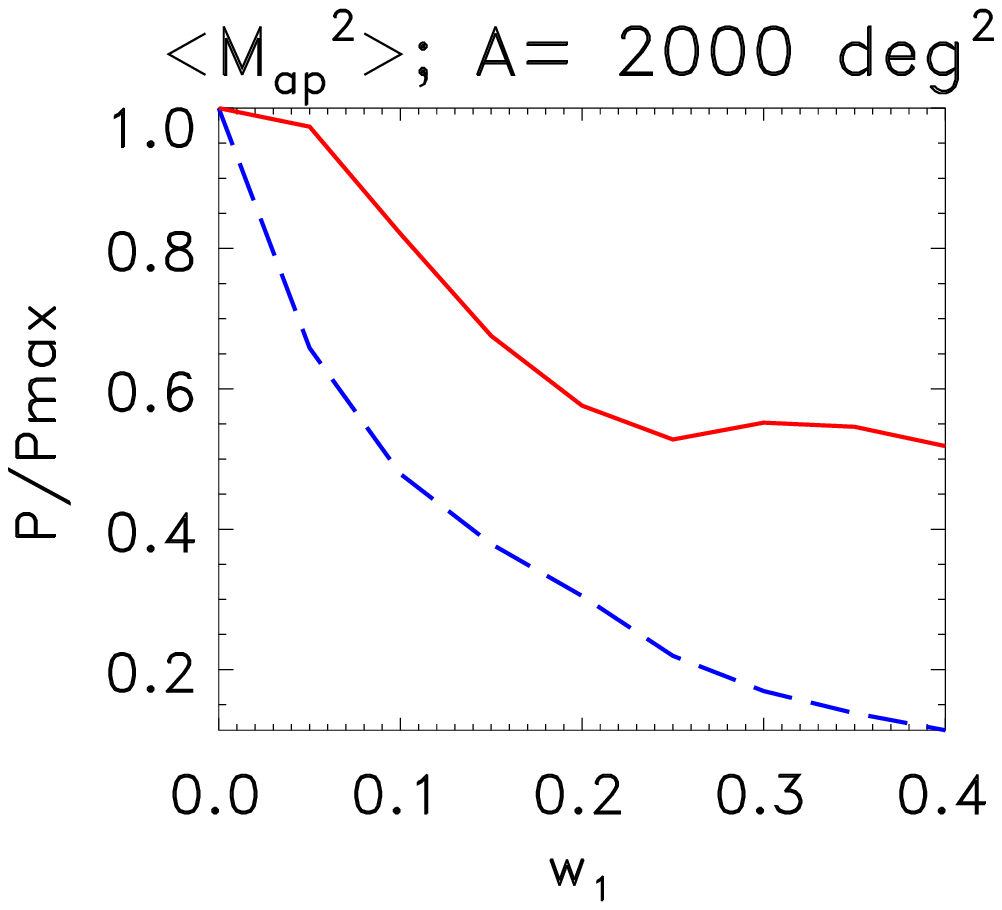,width=.45\textwidth}
\end{tabular}
\caption{\label{res_dark} Cosmological constraints on the parameters of the
  dark energy equation of state, for the survey of area
  $A=2000~ {\rm deg^2}$. Left panel: 68\% and 95\% confidence regions on the 
$[w_0,w_1]$  plane for the non-bias case (solid, red) and marginalised over the bias parameter space
  (dashed, blue). Right panel: the marginal 1-dimensional probability distribution for $w_1$
  obtained assuming no bias (solid line, red) and assuming a fixed bias of $(a_m=-0.02\,,\, b_m=0)$
  (dashed line, blue).}
\end{figure*}

Regarding the impact of the bias, we find the
multiplicative redshift-dependent bias essentially does not affect the non-tomographic
constraints, even for a KIDS-like survey. Indeed, the FoM of the $[\Om,w_0]$
contour only decreases by a factor of 1.1 in the presence of bias. For the 
$[\Om,\sigma_8]$ contour the decrease is larger, around 1.6. The tomographic
constraints are however are however more strongly  affected by the bias. We find  using
tomography  the $[\Om,\sigma_8]$ marginalised FoM is  $\sim 3.5$  smaller  in
comparison to the tomographic  FoM without bias. It is still better than the
bias-affected non-tomographic one by a factor of around 2, but it is only slightly
better than  to FoM for the $A=170 ~ {\rm deg^2}$ survey in the case
of no-bias. Notice however that even though the two FoM are similar, the
corresponding one-dimensional constraints for $\Om$ and $\sigma_8$ are
different. Indeed the two contours (the solid contour in the right panel of
Fig.~\ref{res_wide} and the dashed one in the upper-right panel of
Fig.~\ref{const_res}) have different shapes. This indicates the effect of the
bias is not equivalent to an increase of the data error bars. 
Similarly, marginalising over the bias   causes  the FoM of the  tomographic
$[\Om,w_0]$ constraint to increase  by a factor of 2 in comparison
to the unbiased  FoM.

The finding that tomography is more affected by the bias reflects the fact that  
cosmological and bias parameters are more correlated for this case. We looked
at the bias-cosmology correlations and verified that while for the
non-tomographic case only $\Om$ shows some correlation with $a_m$ and $b_m$,
in the case of tomography  $\Om$, $\sigma_8$ and $w_0$ are all correlated with
$a_m$. In particular, $\sigma_8$ and $a_m$ are anti-correlated, whereas $w_0$
and $a_m$ correlate. This means a lower (more negative) $w_0$ is needed
to compensate the multiplicative bias, which essentially underestimates
the signal.
This result can be explained if one considers how the properties of
large-scale structures depends on  dark energy. For  larger $w_0$, dark
energy starts to dominate at times later than those predicted by 
$\Lambda$CDM, and the structures  grow faster. If one normalises the
amplitude of the density fluctuations to the present time then  models with  $w_0>-1$
 would have less
 structure in the past and   produce  a signal $\langle \map ( \vartheta) \rangle$ which is smaller
 than the signal produced by the $\Lambda$CDM model. The dark energy  affects
 two-point cosmic shear statistics also via the geometric
factor and  this effect has the opposite result increasing the amplitude of $\langle \map ( \vartheta) \rangle$
with $w_0$ \cite{SB05}. This last effect  is dominant  for the correlations
involving the lower redshift bins, whereas the previous effect is dominant
for high redshift bins. 
For our particular choice of redshift bins, the first effect is the
dominant one, as  is shown in Figure~\ref{bias_w0}, by the shift to the left
of the probability distribution, in the presence of bias. This plot also
illustrates that ignoring the presence of a shape measurement bias on the
likelihood analysis may lead to an incorrect cosmological interpretation.

We also performed a likelihood analysis for the $A=2000~{\rm deg^2}$ survey,
 this time using the equation of state of dark
energy given by Eq.~(\ref{sugra}), including  an extra parameter, $w_1$.
The resulting marginalised contours for both the $[\Om,\sigma_8]$ and the
$[\Om,w_0]$ plane are very similar to the results of the previous analysis. 
The left panel of Figure~\ref{res_dark} shows the contours in
the $[w_0,w_1]$ plane.  They are weakly affected by the bias, broadening
mildly towards lower values of $w_0$ consistently with the previous analysis.
$\langle \map ( \vartheta) \rangle$ cannot independently measure $w_1$, as
 shown by the rather flat probability distribution in the right panel of
 Figure~\ref{res_dark} (solid line). However for a bias of
 $(a_m=-0.02\,,\,b_m=0)$, i.e., assuming this bias in the theoretical models
 and keeping the fiducial model bias-free, the probability distribution is
 much narrower. The reason for this is that the impact of this bias on the
 models is much larger than the effect of $w_1$. Therefore, most values of
 $w_1$ cannot compensate for the bias impact. Conversely, when doing a likelihood analysis of 
real data biased by an unknown amount, assuming no bias
 in the theoretical models, the same effect may occur leading to an inaccurate but apparently
 precise estimate.

\section{Conclusions}\label{conclusions}

We have investigated the impact of realistic
 redshift-dependent measurement bias on the estimation of cosmological
 parameters using two-point shear statistics in a tomographic approach with
 three redshift bins.
We focused on two survey types of the same depth, with median redshift $\sim
 0.8$, covering an area of 170
 $\rm {deg}^2$ and 2000 $\rm {deg}^2$, representing current and
 near-future ground-based weak lensing surveys, respectively.

We performed a likelihood  analysis in a grid of cosmological and bias
 parameters. The covariance matrix of the tomographic two-point functions was
 derived analytically in the real space, extending the formulae of Schneider
 et al. (2002), and was afterwards calibrated for non-Gaussianity, 
extending the fitting formula of Semboloni et al. (2007).

Realistic redshift-dependent multiplicative bias were obtained 
 reanalyzing the STEP2 set of simulations with a KSB pipeline.
 The results of our analysis are thus strictly valid for KSB-based
 measurements only. In this framework, our results are also a worst-case
 scenario, since the higher end of the bias values found correspond to the
 highly elliptical PSFs D and E of STEP2, which will be avoided by design in 
 future surveys. Nevertheless, such ellipticities are comparable with those
 measured in the outer regions of CFHTLS-Wide fields. 
 
The top-hat variance is more sensitive to the presence of
 a shape measurement bias than the variance of the aperture
 mass. In particular, in the case of a large scale-independent additive bias,
 cosmological interpretation of the signal may even not be possible, 
 suggesting that aperture mass statistics should  be preferred to top-hat statistics.
 The measurement of  $\langle \map(\vartheta) \rangle$ is however also quite 
sensitive to the shape measurement bias; its amplitude  is reduced by  $20\%$
 at redshift $\sim 1$, and its scaling with redshift becomes shallower.

 We stress the results shown in this paper are based only on the
 simulations. Using these simulations the aperture mass variance  seems to be
 more accurate than the top-hat variance, this because the top-variance is
 particularly sensitive to the additive bias as long as this bias is constant
 across the field. On real data this is not often true and we expect both  two-point
 shear statistics to be biased. Unfortunately, we could not quantify this
 effect using this set of simulations as the PSF is constant across the field.
Moreover, in real data, the profile of the PSF is not constant in time, space and color,
 and it might have a more complex shape. These factors may introduce other
sources of bias which are not investigated here. Finally, in real data,
 the lack of accuracy on the estimation of photometric redshifts could further 
decrease the scaling of the two-point shear statistics with redshift. 
Indeed, the presence of outliers in photometric redshift estimation
 is known to decrease the measured slope of the evolution of the cosmic shear
 signal \cite{Cars08}.

We found the redshift-dependent bias to have, in general, a large impact on
the tomographic measurements. For example, constraints obtained from the
largest survey, marginalised over the realistic bias interval, are comparable
to the ones from the ten times smaller survey assuming the latter does not
suffer from measurement bias.

 Ignoring the presence of a shape measurement bias can already bias 
 the estimation of  cosmological parameters in surveys such as the
 CFHTLS-Wide. From our analysis, the bias on $\sigma_8$ is of a few percent.
 For the survey of $2000~ {\rm deg^2}$ the effect of ignoring the bias is more
 important, around $5\%$, corresponding roughly to $1~\sigma$, for $w_0$, and
 larger for $w_1$.

 The safer approach when dealing with unknown biases is to marginalise over
 a realistic range. While this reduces the constraining power of the tomographic
 analysis over a non-tomographic one, there is the possibility for
 self-calibrating by combining information from other cosmic shear measures.
 However, the best approach would be to correct for these biases. 
 For such a reason, collaborations that aim to improve the measurement of the
 galaxy shapes such as STEP and GREAT08 \cite{Bretal08} are important
 for the  development and  improvement of  shear measurement methods. 
As an example, the simulations produced by the STEP collaboration
have been used to develop and test new PSF-correction methods, such as shapelets-based
decompositions \cite{Ku06} and the LENSFIT method \cite{Kietal08b}, which are 
among the most promising methods for future weak lensing analysing.

To conclude, this analysis, which is limitated to the KSB PSF correction
 method  and uses  only the STEP2 set of simulations shows the importance  to investigate further  the dependence of  the shear measurement bias on the
simulated galaxy properties  in order to infer more accurate constraints  on cosmological
 parameters. Furthermore, simulations cannot be
 used to investigate  all
 the sources of systematic errors  which may be present on real data; for this reason
 one needs  to complement this study using consistency tests with real
 data, for example  measuring the cosmic shear signal  of various galaxy
 populations with  different color, size, magnitude and ellipticity.

\section{Acknowledgments}
We warmly thank the participants of the STEP  collaboration, in particular we
would like to thank  Richard Massey for producing
and making available the set of simulations used for this paper. We also thank the participants of the 
CFHTLS Systematics  collaboration  for helpful comments.
We are grateful to Peter Schneider and Thomas Erben for a careful reading of the manuscript.  
ES acknowledges  the support of  the Alexander von Humboldt  foundation. IT acknowledges the
support of the European Commission Programme 6-th framework, Marie Curie
Training and Research Network ``DUEL'', contract number CT-2006-036133.

\onecolumn
\appendix
\section{Covariance of the tomographic estimator of $\xi_+$}

We extend the calculations of Schneider et al. (2002) (S02 hereafter) in order to explicitly include the redshift binning of the correlation function. 
Note latin subscripts $(i,j)$ refer to individual galaxies, while greek subscripts $(\alpha,\beta)$ refer to bins.

The shear $\gamma$ at angular position $\vc\theta_i$ of a source galaxy at
redshift $z_i$ is estimated from the observed ellipticity of the
galaxy image. After deconvolving the PSF, the corrected ellipticity  $\eps_i$
is related to the intrinsic ellipticity
$\eps^{\rm s}_i$ and the shear by 
\be
\eps_i=\eps^{\rm s}_i+\gamma(\vc\theta_i,z_i)\;.
\label{shearest}
\ee

The correlation function $\xi_+(\vt,\Z(\alpha,\beta))$ between redshift bins with
source redshift distributions $p_\alpha(z)$ and $p_\beta(z)$ is estimated in
angular-separation bins $\vt$.
 The angular bins are defined by
the function $\Delta_\vt(\abs{\vc\theta_i-\vc\theta_j})=1$ for 
$\vt-\Delta\vt/2<\abs{\vc\theta_i-\vc\theta_j}\le\vt+\Delta\vt/2$ and zero otherwise, while the redshift
bins are defined by $\Delta_\Z(z_i,z_j)=1$ for
$p_\alpha(z_i)\times p_\beta(z_j)\neq 0$ and zero otherwise. 
In other words, only pairs of galaxies with one of the galaxies belonging to the bin $\alpha$
and the other to the bin $\beta$ contribute to the correlation at $\Z(\alpha,\beta)$.
The estimator used for the
correlation function is,
\be
\hat \xi_+(\vt,{\cal Z})={ \sum_{ij} w_i\,w_j\,(\eps_{i{\rm +}}\eps_{j{\rm +}}
+ \eps_{i\times}\eps_{j\times})\,
\Delta_\vt(\abs{\vc\theta_i-\vc\theta_j})\Delta_\Z(z_i,z_j)
\over
N_{\rm p}(\vt,\Z)}\;,
\label{xipestimator}
\ee
where the subscripts $(+,\times)$ refer to the two components of the ellipticity, $N_{\rm p}(\vt,\Z)$ is the effective number of pairs contributing
to the correlation at $(\vt,\Z)$ and $w_i$, $w_j$ allow for weighting the galaxies. 
In the absence
of intrinsic alignments and shear-shape correlations,
Eq.~(\ref{xipestimator}) is an unbiased estimator of $\xi_+$. The noise of
the estimator is a function of the ellipticity shape noise, $
\ave{\eps^{\rm s}_{i{\rm +}}\eps^{\rm s}_{j{\rm +}}
+ \eps^{\rm s}_{i\times}\eps^{\rm s}_{j\times}}=\sigma_\eps^2\delta_{ij}\,$, which we
assume is independent of redshift. 

The covariance of $\hat \xi_+(\vt,\Z)$ is defined as
\be
{\rm Cov}(\hat\xi_+,\vt_1,\Z_1;\hat\xi_+,\vt_2,\Z_2)
=\ave{\hat \xi_+(\vt_1,\Z_1)\,\hat \xi_+(\vt_2,\Z_2)}
-\xi_+(\vt_1,\Z_1)\,\xi_+(\vt_2,\Z_2)\;.
\ee
Inserting Eq.~(\ref{xipestimator}) it reads,
\bea
{\rm Cov}(\hat\xi_+,\vt_1,\Z_1;\hat\xi_+,\vt_2,\Z_2)
&=&
\sum_{ijkl}w_iw_jw_kw_l
\Delta_{\vt_1}(ij) \Delta_{\vt_2}(kl)
\Delta_{\Z_1}(ij) \Delta_{\Z_2}(kl) 
\ave{\rund{\eps_{i1}\eps_{j1}+\eps_{i2}\eps_{j2}} 
 \rund{\eps_{k1}\eps_{l1}+\eps_{k2}\eps_{l2} } }
\times
\nonumber \\
&\times& {1 \over N_{\rm p}(\vt_1,\Z_1)N_{\rm p}(\vt_2,\Z_2) }
- \xi_+(\vt_1,\Z_1)\,\xi_+(\vt_2,\Z_2)  \;.
\eea
\\
This expression depends on four-point correlations involving the four positions
$(\theta_i,z_i)$, $(\theta_j,z_j)$, 
$(\theta_k,z_k)$ and $(\theta_l,z_l)$,
 where the first two positions define $(\vt_1,\Z_1)$ and the latter two define 
$(\vt_2,\Z_2)$. In order to evaluate it, we proceed as in S02: inserting Eq.~(\ref{shearest}); assuming a Gaussian shear field and thus factoring four-point functions as a
 sum over products of two-point functions (assuming no intrinsic alignments),
 and finally writing the two-point correlations of the
 shear components as function of $\xi_+$ and $\xi_-$.
 We obtain the results (A5)-(A7), which differ from Eq.~(23) of S02
 since we can no longer interchange $i$ with $j$ or  $k$ with $l$, which may
 have different redshifts.  The cosmic variance term reads
\bea
{\rm V_{++}}(\hat\xi_+,1;\hat\xi_+,2)&=&
{1\over N_{\rm p}(\vt_1,\Z_1)N_{\rm p}(\vt_2,\Z_2)}
{1\over 2}\,\sum_{ijkl}w_iw_jw_kw_l
\Delta_{\vt_1}(ij) \Delta_{\vt_2}(kl) \Delta_{\Z_1}(ij)
\Delta_{\Z_2}(kl)
\Biggl[  \xi_+(il)\xi_+(jk)+
\nonumber \\
&+& \xi_+(ik)\xi_+(jl) 
+ \cos\eck{4\rund{\vp_{il}-\vp_{jk}}}\xi_-(il)\xi_-(jk)+
\cos\eck{4\rund{\vp_{ik}-\vp_{jl}}}\xi_-(ik)\xi_-(jl)
\Biggr]\;,
\label{vapp}
\eea
where $\vp_{ij}$ is the polar angle of the vector $\vc\theta_i-\vc\theta_j$.
The mixed term, with the coupling between shape-noise and shear, is now 
\bea
{\rm M_{++}}(\hat\xi_+,1;\hat\xi_+,2)&=&
{1\over N_{\rm p}(\vt_1,\Z_1)N_{\rm p}(\vt_2,\Z_2)}
{\sigma_\eps^2 \over 2} \Biggl [ 
\sum_{ijk}w_i^2w_jw_k 
\Delta_{\vt_1}(ij) \Delta_{\vt_2}(ik) \Delta_{\Z_1}(ij) \Delta_{\Z_2}(ik)
\xi_+(jk) + 
\nonumber \\
&+& 
\sum_{ijk}w_iw_j^2w_k 
\Delta_{\vt_1}(ij) \Delta_{\vt_2}(jk) \Delta_{\Z_1}(ij) \Delta_{\Z_2}(jk)
\xi_+(ik)+
\sum_{ijl}
w_i^2w_jw_l
\Delta_{\vt_1}(ij) \Delta_{\vt_2}(il) \Delta_{\Z_1}(ij) \Delta_{\Z_2}(il)
\xi_+(jl) + 
\nonumber \\
&+&
\sum_{ijl} w_iw_j^2w_l 
\Delta_{\vt_1}(ij) \Delta_{\vt_2}(jl) \Delta_{\Z_1}(ij) \Delta_{\Z_2}(jl)
\xi_+(il)
\Biggr ] \;.
\label{mapp}
\eea
The noise term is similar to the non-tomographic case, only contributing to the
 diagonals of the covariance matrix, i.e. for $\vt_1=\vt_2$ and $\Z_1=\Z_2$. It reads,
\be
{\rm S}\,(\hat\xi_+,1;\hat\xi_+,2)=
{1\over N_{\rm p}(\vt_1,\Z_1)N_{\rm p}(\vt_2,\Z_2)}
\sigma_\eps^4  \sum_{ij} w_i^2w_j^2 \Delta_{\vt_1}(ij)
 \Delta_{\vt_2}(ij) \Delta_{\Z_1}(ij) \Delta_{\Z_2}(ij) \;. 
\label{sapp}
\ee
\\
Table \ref{tabapp} summarizes the dependence of the cosmic variance on the
tomographic auto- and cross-correlation functions, considering two 
redshift bins for illustration purposes. In the case the cross-correlation signal is not considered in the data vector,
 the matrix reduces to the sub-matrix highlighted, which nonetheless depends on the
 tomographic cross-correlation. Note this restricted case
corresponds to imposing $p_\alpha(z)=p_\beta(z)$ being thus invariant under $i-j$ and
$k-l$ transformations, and therefore Eqs.~(\ref{vapp})-(\ref{sapp}) reduce to
Eq.~(23) of S02.

\begin{table}
\centering
\label{tabapp}
\begin{tabular}{c|ccrc}
 & $\xi_{11}$ & $\xi_{22}$ & & $\xi_{12}$ \\
\hline
$\xi_{11}$  & $\xi_{11}\,\xi_{11}$ + S & $\xi_{12}\,\xi_{12}$ & $|$ & $\xi_{11}\,\xi_{12}$ \\
$\xi_{22}$  & $\xi_{12}\,\xi_{12}$ & $\xi_{22}\,\xi_{22}$ + S & $|$ & $\xi_{22}\,\xi_{12}$ \\
& ||| & ||| &  & \\
$\xi_{12}$  & $\xi_{11}\,\xi_{12}$ & $\xi_{22}\,\xi_{12}$ & & $\xi_{11}\,\xi_{22}+\xi_{12}\,\xi_{12}$ + S\\
\hline
\end{tabular}
\caption{Summary of the tomographic cosmic variance matrix, with dimension $3\times 3$
  corresponding to 2 redshift bins $(\alpha=[1,2])$, showing the dependence on the various correlation functions 
$\xi_+(\vt,\Z(\alpha,\beta))$ (denoted by $\xi_{\alpha \beta}$). The terms affected by shot noise are also indicated (with S).
 Dashed lines isolate the restricted case of neglecting the information in the cross-correlation of redshift bins.}
\end{table}

The calculations in Eqs.~(\ref{vapp})-(\ref{sapp}) are time-consuming and
 survey-dependent since
 they involve sums over the galaxies positions. Still following S02,
 we calculate a more convenient quantity: the ensemble average
 of the covariance matrix for a survey of area $A$ and galaxy density $n$, and write the result in the form
\be
{\rm E}\rund{{\rm Cov(\,\hat\xi_+,\vt_1,\Z_1(\alpha_1,\beta_1);\hat\xi_+,\vt_2,\Z_2(\alpha_2,\beta_2)\,)}}
=D\,\delta_{\rm D}(\vt_1-\vt_2)\,\delta_{\rm D}(\Z_1-\Z_2)+q_{++}+r_{+0}+r_{+1}.
\label{averagecov}
\ee
For the cosmic variance we obtain
\bea\label{cosmic_var_term1}
r_{+0}&=&{1\over \pi A}\int_0^\infty\d\phi\,\phi \left[
\int_0^\pi\,\d\vp_1\,\xi_{+,\alpha_1\alpha_2}(|\vc\psi_a|)
\int_0^\pi\,\d\vp_2\,\xi_{+,\beta_1\beta_2}(|\vc\psi_b|)+
\int_0^\pi\,\d\vp_1\,\xi_{+,\alpha_1\beta_2}(|\vc\psi_a|)
\int_0^\pi\,\d\vp_2\,\xi_{+,\beta_1\alpha_2}(|\vc\psi_b|)
\right] \;,
\\
r_{+1}&=&{1\over (4\pi) A}\int_0^\infty\d\phi\,\phi
\int_0^{2\pi}\d\vp_1\,\xi_{-,\alpha_1\alpha_2}(|\vc\psi_a|)
\int_0^{2\pi}\d\vp_2\,\xi_{-,\beta_1\beta_2}(|\vc\psi_b|)
\eck{\cos 4\vp_a \,\cos 4\vp_b + \sin 4\vp_a \,\sin 4\vp_b}
\label{cosmic_var_term2}\\
&+&
{1\over (4\pi) A}\int_0^\infty\d\phi\,\phi
\int_0^{2\pi}\d\vp_1\,\xi_{-,\alpha_1\beta_2}(|\vc\psi_a|)
\int_0^{2\pi}\d\vp_2\,\xi_{-,\beta_1\alpha_2}(|\vc\psi_b|)
\eck{\cos 4\vp_a \,\cos 4\vp_b + \sin 4\vp_a \,\sin 4\vp_b}\;\nonumber,
\eea
where $\xi_{+,\alpha\beta}$ is a short-hand notation for $\xi_+(\vt,\Z(\alpha,\beta))$.
The result was obtained as in S02: the integrations arise from the
averaging operator, and a change of variable is made from the position angles $\vc\theta$, included in
 Eq.~(\ref{vapp}), to separation angles $\vc\phi$. We note that due to the loss
 of the $i-j, \, k-l$ invariances, different changes of
 variables are needed for the two terms of $r_{+0}$ or $r_{+1}\,;$ $\vc\phi$ is defined as
 $\vc\phi= \vc\theta_3-\vc\theta_1$ in one case and $\vc\phi=
 \vc\theta_3-\vc\theta_2$ in the other.
The angular separation information is contained in the vectors
 $\vc\psi_a$, $\vc\psi_b$ and their polar angles $\vp_a$ and $\vp_b$:
\be
\vc\psi_a=\vectii{\phi\cos\vp-\vt_1\cos\vp_1}{\phi\sin\vp-\vt_1\sin\vp_1}
\quad ;\quad
\vc\psi_b=\vectii{\phi\cos\vp+\vt_2\cos\vp_2}{\phi\sin\vp+\vt_2\sin\vp_2}
\ee
\\
For the mixed term we find
\bea\label{mixed}
q_{++}&=&{1\over 2 \pi A }
\int_0^\pi\d\vp\;\left[
{ \sigma_{\eps,\beta_1\beta_2}^2\over \sqrt{n_{\alpha_1}n_{\alpha_2}}}\,\xi_{+,\alpha_1\alpha_2}\rund{\sqrt{\vt_1^2+\vt_2^2-2\vt_1\vt_2\cos\vp}}
+ {\sigma_{\eps,\alpha_1\alpha_2}^2\over \sqrt{n_{\beta_1}n_{\beta_2}}}\,\xi_{+,\beta_1\beta_2}\rund{\sqrt{\vt_1^2+\vt_2^2-2\vt_1\vt_2\cos\vp}}
\right. +
\nonumber \\
&+& \left. 
{\sigma_{\eps,\beta_1\alpha_2}^2\over \sqrt{n_{\alpha_1}n_{\beta_2}}}\,\xi_{+,\alpha_1\beta_2}\rund{\sqrt{\vt_1^2+\vt_2^2-2\vt_1\vt_2\cos\vp}}
+ {\sigma_{\eps,\alpha_1\beta_2}^2\over \sqrt{n_{\beta_1}n_{\alpha_2}}}\,\xi_{+,\beta_1\alpha_2}\rund{\sqrt{\vt_1^2+\vt_2^2-2\vt_1\vt_2\cos\vp}}
\right]\,.
\eea
The expectation value of the shot noise term is
\be\label{noise}
D={ 1 \over {A \,4\pi \,\vt \,\Delta\vt} }\left[
{ \sigma_{\eps,\alpha_1\alpha_2}^2\over \sqrt{n_{\alpha_1}n_{\alpha_2}}}\,
{ \sigma_{\eps,\beta_1\beta_2}^2\over \sqrt{n_{\beta_1}n_{\beta_2}}}
+ { \sigma_{\eps,\alpha_1\beta_2}^2\over \sqrt{n_{\alpha_1}n_{\beta_2}}}\,
{ \sigma_{\eps,\beta_1\alpha_2}^2\over \sqrt{n_{\beta_1}n_{\alpha_2}}}
\right], 
\ee
where $\Delta\vt$ is the angular bin size. We assume $\sigma_{\eps,\alpha_1\alpha_2}^2=\sigma_{\eps,\beta_1\beta_2}^2=
\sigma_{\eps,\alpha_1\beta_2}^2=\sigma_{\eps,\beta_1\alpha_2}^2=\sigma_\eps^2$

Finally, we insert Eq.~(\ref{averagecov}) in Eq.~(42) of S02 to obtain the
covariance matrices for the aperture-mass and top-hat dispersion.
In the absence of B-modes they depend only on the covariance of $\xi_+$, hence
we do not calculate ${\rm Cov}(\hat\xi_-;\hat\xi_-)$ or ${\rm Cov}(\hat\xi_+;\hat\xi_-)$.

\end{document}